\documentclass[acmtog]{acmart}

\usepackage{booktabs} %

\usepackage{geometrycollective}
\usepackage{multirow}

\usepackage[draft]{fixme}

\usepackage[ruled]{algorithm2e} %

\SetAlFnt{\small}
\SetAlCapFnt{\small}
\SetAlCapNameFnt{\small}
\SetAlCapHSkip{0pt}
\IncMargin{-\parindent}

\acmJournal{TOG}
\acmVolume{X}
\acmNumber{X}
\acmArticle{XX}
\acmYear{XXXX}
\acmMonth{-1}

\setcopyright{acmcopyright}
\acmDOI{XX}

\received{January 20XX}

\begin{teaserfigure}
  \centering
   \includegraphics[width=\textwidth]{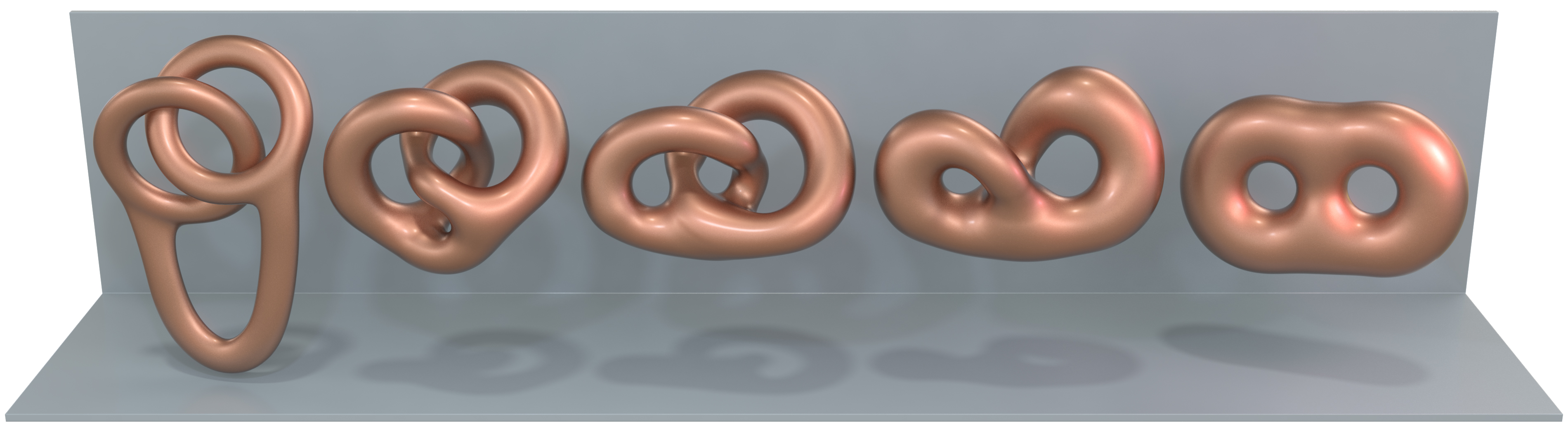}
   \caption{We develop a numerical scheme for optimizing surface geometry while avoiding self-intersections.  Here, we automatically find an unexpected transition between linked and unlinked states of a pair of ``handcuffs'' by simply minimizing a repulsive energy.\label{fig:Teaser}}
\end{teaserfigure}

\definecolor{forestGreen}{RGB}{34, 139, 34}

\newcommand{\AmbDim}{m}
\newcommand{\DomDim}{n}
\newcommand{\ConstraintDim}{k}

\newcommand{\abs}[1]{\left\lvert#1\right\rvert} %
\newcommand{\nabs}[1]{\lvert{#1}\rvert} %

\newcommand{\ninnerprod}[1]{\langle #1 \rangle}
\newcommand{\innerprod}[1]{\left\langle #1 \right\rangle}
\newcommand{\Domain}{M} %

\newcommand{\Ecal}{\mathcal{E}}

\newcommand{\Ical}{{\mathcal{I}}}
\newcommand{\Jcal}{{\mathcal{J}}}
\newcommand{\Df}{\mathcal{D}_f}
\newcommand{\DMatrix}{\mathsf{D}_f}

\newcommand{\Energy}{\Ecal}
\newcommand{\EnergyP}{\Ecal^{p}}
\newcommand{\DiscEnergy}{E}
\newcommand{\DiscEnergyP}{\DiscEnergy^{p}}
\newcommand{\ApproxDiscEnergyP}{\widetilde{\DiscEnergy}^{p}}

\newcommand{\DDiscEnergyP}{d \DiscEnergyP}

\newcommand{\ConstraintFn}{\Phi}
\newcommand{\DConstraintFn}{d\ConstraintFn}

\DeclareMathOperator{\diag}{diag}

\providecommand{\T}{\top}

\newcommand{\FaceAv}[1]{\bar{#1}}

\newcommand{\Bary}[1]{X_{#1}}
\newcommand{\Proj}{P}
\newcommand{\PreLo}{\mathsf{U}}
\newcommand{\PreHi}{\mathsf{V}}
\newcommand{\KernelMatrix}{H}
\newcommand{\Kernel}{h}

\begin{document}
\title{Repulsive Surfaces}
\author{Chris Yu}
\affiliation{%
  \institution{Carnegie Mellon University}
  }
\author{Caleb Brakensiek}
\affiliation{%
  \institution{Independent Researcher}
  }
\author{Henrik Schumacher}
\affiliation{%
  \institution{RWTH Aachen University}
  \streetaddress{Templergraben 55}
  \city{Aachen, Germany}
  \postcode{52062}
  }
\author{Keenan Crane}
\affiliation{%
  \institution{Carnegie Mellon University}
  \streetaddress{5000 Forbes Ave}
  \city{Pittsburgh}
  \state{PA}
  \postcode{15213}}

\renewcommand\shortauthors{Chris Yu and Caleb Brakensiek and Henrik Schumacher and Keenan Crane}
\begin{abstract}
   Functionals that penalize bending or stretching of a surface play a key role in geometric and scientific computing, but to date have ignored a very basic requirement: in many situations, surfaces must not pass through themselves or each other.  This paper develops a numerical framework for optimization of surface geometry while avoiding (self-)collision.  The starting point is the \emph{tangent-point energy}, which effectively pushes apart pairs of points that are close in space but distant along the surface.  We develop a discretization of this energy for triangle meshes, and introduce a novel acceleration scheme based on a fractional Sobolev inner product.  In contrast to similar schemes developed for curves, we avoid the complexity of building a multiresolution mesh hierarchy by decomposing our preconditioner into two ordinary Poisson equations, plus forward application of a fractional differential operator.  We further accelerate this scheme via hierarchical approximation, and describe how to incorporate a variety of constraints (on area, volume, \etc{}).  Finally, we explore how this machinery might be applied to problems in mathematical visualization, geometric modeling, and geometry processing.
\end{abstract}

\ccsdesc[500]{Computing methodologies~Shape modeling}
\ccsdesc[500]{Mathematics of computing~Continuous optimization}

\keywords{Computational design, shape optimization, surfaces}

\thanks{This work is supported by the National Science Foundation, under grant XXXX, grant XXXX and grant XXXX.}

\maketitle

\section{Introduction and Related Work}
\label{sec:Introduction}

A geometric functional assigns a real-valued score \(\mathcal{E}(f)\) to each immersion \(f: M \to \mathbb{R}^\AmbDim\) of a surface \(M\).  Such functionals serve as regularizers in many geometric problems, helping to define a unique solution, or simply making the geometry ``nicer'' in some sense.  For instance, in geometric modeling they are used to smoothly interpolate given boundary data~\cite{Bucur:2006:VMS}, in mathematical visualization they can be used to endow an abstract surface with a concrete geometry~\cite{Chern:2018:SFM}, and in digital geometry processing they are used for, \eg{}, hole filling~\cite{Clarenz:2004:FEM} or denoising of measured data~\cite{Elsey:2009:ATV}.  However, classic functionals ignore a basic requirement of many applications---namely, that surfaces should not exhibit \hbox{(self-)}intersections.  This condition is critical when surfaces represent physical membranes (\eg{}, in biological simulation), boundaries of solid objects (\eg{}, for digital manufacturing), or certain mathematical objects (\eg{}, isotopy classes of embeddings).  It is therefore surprising that, to date, there has been little focus on interpenetration in variational surface modeling.  We build on the recent framework of \citet{Yu:2021:TOG}, extending their machinery for repulsive curves to the more computationally demanding case of surfaces.

\paragraph{Curvature Functionals.} A basic functional for surfaces is total surface area; gradient descent on total area leads to \emph{mean curvature flow}, which has been used for surface denoising~\cite{Desbrun:1999:IFI} but can develop non-smooth singularities or pinch-off artifacts.  Though efforts have been made to desingularize this flow~\cite{Kazhdan:2012:CMC}, sharp peaks and cusps are ultimately impossible to detect from area alone.  For this reason, functionals used in geometric modeling typically incorporate curvature information---most prominently the \emph{Willmore energy} \(\mathcal{E}_W(f) := \int_M (H^2 - K) \ dA\), where \(H\) and \(K\) are the mean and Gaussian curvatures, \resp{}\  Significant work has focused on numerical optimization of Willmore energy~\cite{Droske:2004:LSF,Bobenko:2005:DWF,Crane:2013:RFC,Soliman:2021:CWS}, but since this energy is M\"{o}bius invariant, it effectively provides a notion of regularity for surfaces in the 3-sphere \(S^3\), rather than Euclidean \(\mathbb{R}^3\).  In the context of geometric modeling, this means that even minimizers of Willmore energy can have poor distributions of curvature---see for example \figref{WillmoreVsTangentPoint}, \figloc{bottom left}.  Though further energies have been developed to address such issues~\cite{Moreton:1992:FOF,Joshi:2007:EMC}, none of these energies avoid intersections.

\setlength{\columnsep}{1em}
\setlength{\intextsep}{0em}
\begin{wrapfigure}{l}{115pt}
   \includegraphics{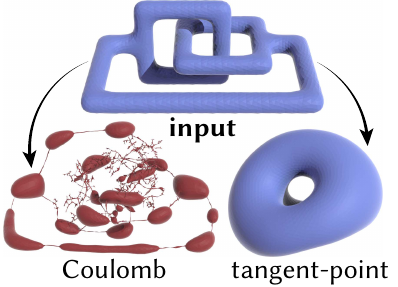}
   \caption{Ad-hoc schemes such as vertex-vertex Coulomb forces do not correspond to a meaningful smooth energy, and can be numerically unstable.  Here we minimize Coulomb and tangent-point energies subject to a fixed area constraint.\label{fig:Coulomb}}
\end{wrapfigure}
\paragraph{Repulsive Forces.}  Collision response forces from physical simulation~\cite{Bridson:2002:RTC} and contact mechanics~\cite{Wriggers:2004:CCM} can be used to locally \emph{resolve} contact, but do not help to guide shape optimization toward a state that is far from interpenetration.  Moreover, whereas level set representations of geometry ensure (by construction) that surfaces have no self-intersections, the \emph{raison d'\^{e}tre} of such methods is to allow the surface topology to \emph{change}, rather than to preserve it~\cite{Osher:2006:LSM}.  We instead consider ``all-pairs'' energies of the form
\[
   \mathcal{E}(f) = \int_{M \times M} k(x,y)\ dx_f \ dy_f,
\]
where \(dx_f\) denotes the area element induced by \(f\), and the kernel \(k: M \times M \to \mathbb{R}\) is designed to discourage self-contact.

A tempting choice is a Coulomb-like potential
\[
   k_{\text{Coulomb}}(x,y) = \frac{1}{|f(x)-f(y)|^\alpha}
\]
for some falloff parameter \(\alpha > 0\); on a triangle mesh, this amounts to just penalizing the distance between all pairs of vertices.  However, as noted by \citet[Section 3.1]{Yu:2021:TOG}, the resulting energy is too weak to prevent collision for \(\alpha < 2\), and yet ill-defined in the continuum limit for \(\alpha \geq 1\).  The essential difficulty is that there are \emph{always} points \(y\) within an arbitrarily small geodesic distance \(d(x,y) \geq |f(x)-f(y)|\) of any point \(x \in M\) \emph{along} the surface, causing the energy to blow up.  Numerically, ad-hoc vertex-vertex penalties are hence unstable and highly unpredictable (\figref{Coulomb}).

For curves, the \emph{M\"{o}bius energy}~\cite{OHara:1991:EK} regularizes the Coulomb potential by subtracting the contribution of points that are nearby on the surface:
\[
	k_{\text{M\"{o}bius}}(x,y) = \frac{1}{|f(x)-f(y)|^2} - \frac{1}{d(x,y)^2}.
\]
This energy is well-defined and strong enough to prevent collisions (for suitable \(\alpha\)), but has two significant drawbacks for geometric modeling.  First, like Willmore energy, M\"{o}bius energy is invariant to M\"{o}bius transformations---leading in this case not only to uneven curvature (\cite[Figure 5]{Kusner:1998:MIK}), but also ``tight spots'' where points distant in \(S^3\) become arbitrarily close when projected into \(\mathbb{R}^3\) (see \cite[Figure 3]{Yu:2021:TOG}).  Second, the geodesic distance \(d(x,y)\), though easy to compute for curves, is prohibitively expensive to compute for all pairs of points on a surface---much less to differentiate with respect to motions of the surface.

\begin{figure}[t]
   \includegraphics{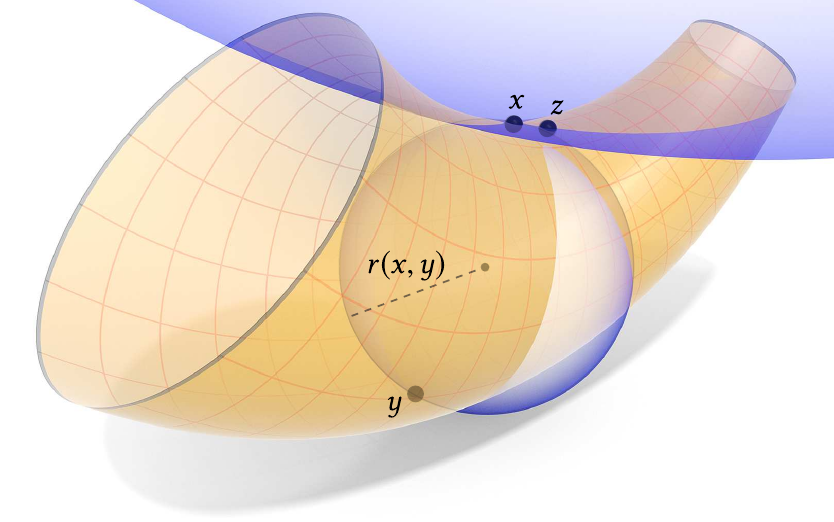}
   \vspace{-2\baselineskip}\caption{For each pair of points \(x,y\) on the surface, the tangent-point energy considers the radius \(r(x,y)\) of the smallest sphere tangent to \(x\) and passing through \(y\), penalizing \(1/r(x,y)\).  Hence, the contribution will be very large for points \(y\) close in space but distant along the surface---and small for points \(z\) nearby along the surface, where the radius is huge.\label{fig:SurfaceTangentPoint}}
\end{figure}

\paragraph{Tangent-Point Energy.} For all these reasons, we are prompted to instead consider the \emph{tangent-point energy} introduced for curves by \cite{Buck:1995:ASE} and extended to higher dimensions by \cite{Strzelecki:2013:JGA}.  For each pair of points \(x,y \in M\), this energy considers the radius \(r(x,y)\) of the smallest sphere tangent to \(f(x)\) and passing through \(f(y)\) (\figref{SurfaceTangentPoint}). The kernel \(k\) is then proportional to \(1/r(x,y)\); \secref{TangentPointEnergy} and \cite{MR3915938} provide further discussion.  Hence, points that are close in space but distant along the surface are penalized; points that are close in space only because they are also close along the surface are ignored.  This energy has several features that make it a prime candidate for repulsive surface optimization, namely:
\begin{itemize}
   \item It provides an infinite barrier to self-intersection~\cite{Strzelecki:2013:JGA}.
   \item Like Willmore energy it penalizes bending~\cite[Section 3.2]{Yu:2021:TOG}, preventing singularities and cusps.
   \item Unlike Willmore and M\"{o}bius energy it is neither M\"{o}bius nor scale invariant, helping to evenly distribute curvature and avoid tight spots.
   \item Unlike M\"{o}bius energy it does not require geodesic distances, and instead depends only on quantities like surface normals \(N\) and extrinsic distances \(|f(x)-f(y)|\) that are cheap to compute and easy to differentiate.
\end{itemize}
However, there are still two significant challenges in applying tan\-gent-point energy to practical surface optimization, namely, (i) picking an inner product that accelerates optimization and (ii) efficiently inverting this inner product.

\paragraph{Accelerating Optimization.} To integrate a parabolic gradient flow of order \(k\) with average node spacing \(h\), one must typically take time steps of size around \(O(1/h^k)\), which is prohibitively expensive for fine meshes.  However, one can effectively transform gradient descent into a 0th-order equation by defining the gradient with respect to a different inner product---mitigating the time step restriction.  This idea of \emph{Sobolev gradients} has long been applied to surface flows~\cite{Pinkall:1993:CDM,Renka:1995:MSS,Eckstein:2007:GSF,Martin:2013:ENO,Schumacher:2017:HGF,Soliman:2021:CWS}, and more recently to elastic energies in geometry processing~\cite{Kovalsky:2016,Claici:2017:IAP,Zhu:2018}.  However, all this work considers energies with integer-order differentials, whereas the tangent point energy has a differential of \emph{fractional} order.  As recently demonstrated by \citet{Yu:2021:TOG}, a fractional inner product hence performs far better than even integer Sobolev schemes, especially for finely-tessellated or highly-knotted curves.  We adopt the same basic strategy, adapting it to surfaces.

\paragraph{Efficient Evaluation.} A second challenge is that there is a dramatic increase in problem size when going from curves to surfaces: rather than integrate an energy over all \(O(n^2)\) pairs of elements on a curve, we now must consider \(O(n^4)\) element pairs on a surface (where \(n \approx 1/h\)).  Standard hierarchical Barnes-Hut approximation is still sufficient to approximate the energy and its differential (\secref{BarnesHut}), but we must also invert the fractional Sobolev inner product, which is now a dense matrix with \(O(n^4)\) entires.  \citet{Yu:2021:TOG} use a multigrid solver based on a simple multiresolution curve hierarchy, but building a multiresolution \emph{surface} mesh hierarchy on each optimization step is far more difficult and expensive.  Our key insight is that the inverse of our fractional operator can be approximated by the inverse of two ordinary (integer-order) Laplace operators, together with \emph{forward} application of a lower-order fractional derivative (\secref{Preconditioner}).  Since this decomposition is only approximate in the discrete setting, we use it to precondition an iterative linear solver (GMRES) that does not require a mesh hierarchy.

Overall our acceleration strategy leads to a straightforward implementation that still provides acceleration sufficient to handle the  challenging surface case.  To give a rough sense of performance, using four threads it takes about 1--2 seconds per descent step on a mesh of about 30,000 triangles, which we have found suitable for interactive work (especially since each step makes considerable progress relative to ordinary gradient descent---see \figref{BumpyMug}).

\begin{figure}
   \includegraphics[width=\columnwidth]{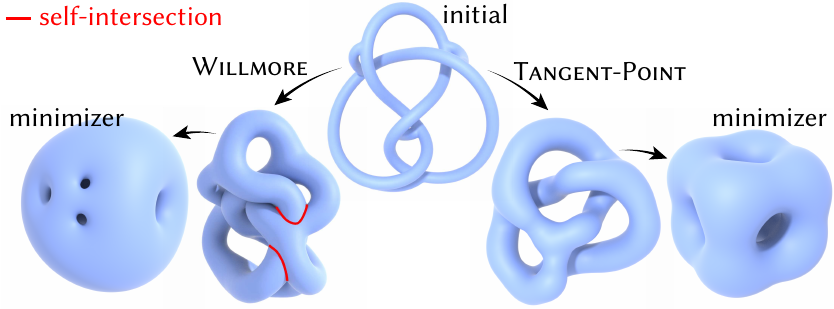}
   \vspace{-2\baselineskip}\caption{Willmore energy does nothing to prevent intersections (in red), and can have minimizers that asymmetrically distribute curvature over the surface.  \figloc{Right:} tangent-point energy avoids intersections and tends to provide a more uniform curvature distribution.\label{fig:WillmoreVsTangentPoint}}\vspace{-\baselineskip}
\end{figure}

\subsection{Contributions}
\label{sec:Contributions}

Overall, in this paper we develop
\begin{itemize}
   \item the first discretization of tangent-point energy for surfaces,
   \item a novel preconditioner that avoids a multigrid hierarchy,
   \item a hierarchical solver that scales to large meshes, and
   \item a framework for handling auxiliary constraints and penalties.
\end{itemize}
We also perform a preliminary investigation of applications in geometric modeling, mathematical visualization, and geometry processing.  Notably, although one can prove that minimizers of the tangent-point energy exist~\cite[Theorem~2]{1504.04538}, these proofs are non-constructive.  Since we provide the first discretization and optimization procedure for the tangent-point energy on surfaces, we obtain the very first glimpse (experimentally) at what some of these surfaces might actually look like.

We begin by defining our problem in the smooth setting (\secref{SmoothEnergy}), followed by a novel discretization of the tangent-point energy and a basic numerical strategy for minimizing it subject to constraints (\secref{Discretization}). We then significantly accelerate this strategy in two distinct ways.  First, we choose an inner product in the smooth setting that vastly improves the convergence of the gradient flow (\secref{FractionalLaplacian}).  Second, in the discrete setting, we propose a preconditioner that dramatically reduces the cost of solving for the descent step (\secref{PreconditionedIterativeMethod}).  We also accelerate evaluation of the energy and its derivatives, as well as dense matrix-vector products, using hierarchical acceleration (\!\secrefs{BarnesHut,PreconditionedIterativeMethod}). We then consider dynamic remeshing (\secref{DynamicRemeshing}) and auxiliary penalties and constraints (\secref{ConstraintsandPenalties}), which enable a variety of potential applications (\secref{ExamplesandApplications}); \secref{EvaluationandComparisons} provides numerical validation.

\begin{figure*}
   \includegraphics[width=\textwidth]{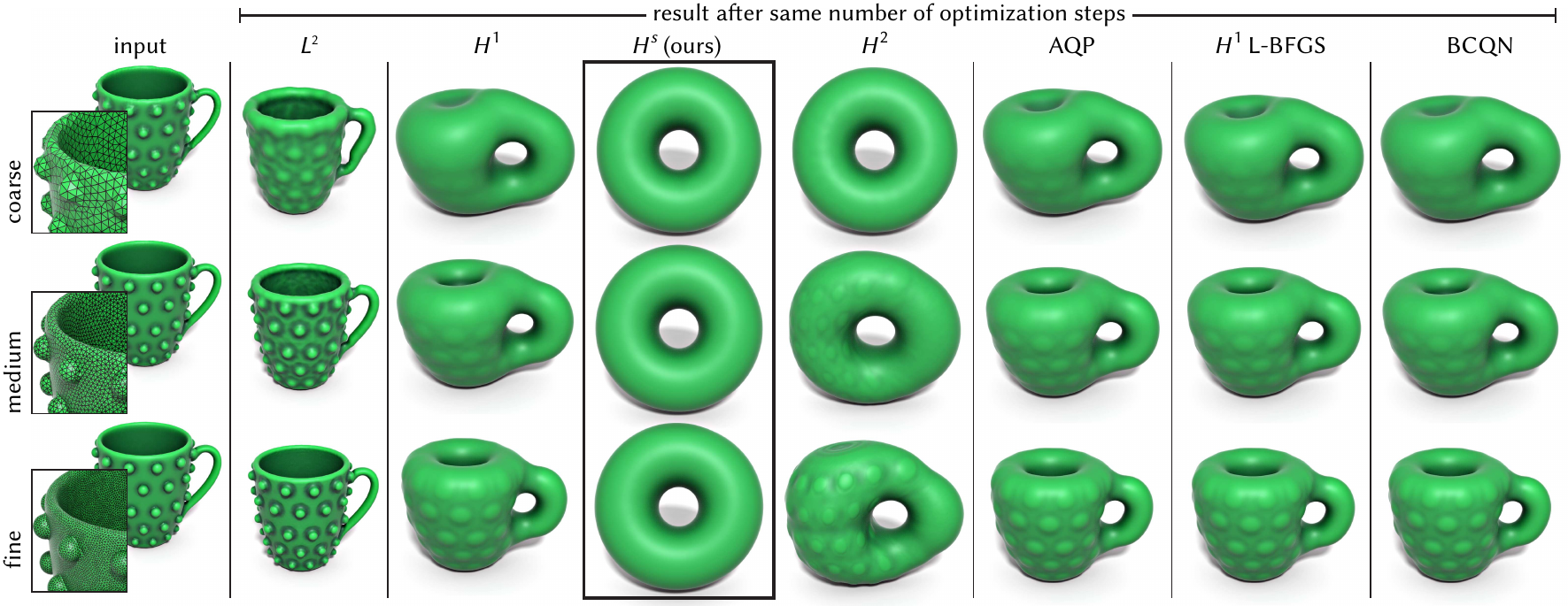}
   \caption{Unlike other schemes, our fractional preconditioner does not suffer from a mesh-dependent time step restriction.  
   Here for example we take 300 optimization steps of maximum size (determined by line search) for each scheme.  
   As resolution increases, all methods but \(H^s\) make slower and slower progress.  
   Note also that schemes based on \(H^1\) preconditioning (\(H^1\), \(H^1\) L-BFGS, AQP, BCQN) quickly eliminate high-frequency details but are slower to smooth the bulk shape; conversely, \(H^2\) quickly smooths out the bulk shape but fine details remain.  
   Using \(H^s\) for \(1 < s < 2\) nicely handles both local and global features.\label{fig:BumpyMug}}
\end{figure*}

\section{Smooth Formulation}
\label{sec:SmoothEnergy}

In this section we define the smooth tangent-point energy \(\EnergyP\), and give some remarks on the order of derivatives appearing in its differential \(d\EnergyP\).  Determining the order of the differential is essential to accelerating the gradient flow \(\tfrac{d}{dt} f = -d\EnergyP(f)\), since it enables us to define a new inner product (in \secref{FractionalLaplacian}) with respect to which the gradient flow effectively becomes a 0th-order equation.  (Readers may wish to consult \citet[Section 4.1]{Yu:2021:TOG} for a slower-paced, didactic introduction to this approach.) Hence, the numerical integrator developed in \secref{Discretization} will be able to take dramatically larger time steps, of a size that does not depend strongly on mesh resolution (\figref{BumpyMug}).

\subsection{Energy}
\label{sec:TangentPointEnergy}

As discussed in \secref{Introduction}, we can define a repulsive energy by considering the \emph{tangent-point radius} $r_f(x, y)$, defined as the radius of the smallest sphere tangent to $f(x)$ and passing through $f(y)$ (\figref{SurfaceTangentPoint}). Letting \(N_f(x)\) be the unit normal at \(x\), this radius can be computed as
\begin{equation}
	\label{eq:TangentPointRadius}
	r_f(x,y) 
	= 
	\frac{\nabs{f(x) - f(y)}^2}{2\nabs{\Proj_f(x)\, (f(x) - f(y)) }},
\end{equation}
where $\Proj_f(x) = N_f(x) \, N_f(x)^{\T}$ denotes orthogonal projector onto the normal space at $x$.  Note that expressing \(r_f\) via the projector avoids picking a sign for the normal, which will be useful in \secref{HierarchicalMatrices} (it is also valid for submanifolds of arbitrary dimension and codimension).  Omitting the constant factor~$2$, the tangent-point kernel (due to \citet{Buck:1995:ASE}) is then given by
\begin{equation}
	k_{f,p}(x,y) := \frac{2^p}{r_f(x,y)^p} 
	= 
	\frac{\nabs{\Proj_f(x)\, (f(x) - f(y)) }^p}{\nabs{f(x) - f(y)}^{2p}}
\end{equation}
for some $p > 0$, and hence the energy itself is
\begin{equation}
\label{eq:TangentPointEnergy}
\EnergyP(f) := \iint_{\Domain^2} k_{f,p}(x,y) \ dx_f \ dy_f .
\end{equation}
While in principle it is possible to allow the exponents in the numerator and denominator to vary independently~\cite{Blatt:2015:CalcVar}, we use exponents $p, 2p$ (as above), which simplifies analysis.
Note that, because $k_{f,p}$ has units $\text{m}^{-p}$ (in meters) and $\EnergyP$ is a double integral over an $\DomDim$-dimensional manifold, $\EnergyP$ has units $\text{m}^{2 \DomDim-p}$.
Therefore, $p >2n$ is required for the energy to be truly repulsive (\ie, to have units corresponding to inverse meters); otherwise, the energy could be reduced to $0$ by simply shrinking the domain to a single point. As we deal with surfaces here ($\DomDim=2$), $p>4$ is sufficient.
Unless otherwise noted, we use $p = 6$ for all examples in this paper.

\subsection{Gradient Flow}
\label{sec:FractionalInnerProduct}

Attempting to perform standard $L^2$ gradient descent on the tangent-point energy yields a flow
\[
\frac{d}{dt}f = -d\Energy^{p}(f).
\]
This flow exhibits poor convergence due to the presence of high-order spatial derivatives on the right-hand side, which even aggressive line search or general-purpose preconditioning (\eg{}, L-BFGS) cannot alleviate; see \figref{BumpyMug}.  However, we can obtain a different descent strategy by defining the gradient with respect to a different inner product.  In particular, if \(A\) is the linear operator defining the inner product, the descent equation becomes 
\begin{equation}
\label{eq:PreconditionedGradientFlow}
\frac{d}{dt}f = - A^{-1} d\Energy^{p}(f).
\end{equation}
An optimal choice of \(A\) will match the order of the differential, so that the right hand side no longer involves any spatial derivatives (hence avoiding a mesh-based time step restriction).  We first establish the order of the differential \(d\EnergyP\) in the surface case (\secref{OrderoftheDifferential}), then define a \emph{fractional Sobolev inner product} that matches this order (\secref{FractionalInnerProduct}).

\subsection{Order of the Differential}
\label{sec:OrderoftheDifferential}

Though originally defined for curves, the tangent-point energy $\EnergyP$ can be formulated for a quite broad class of $\DomDim$-dimensional sets $\varSigma \subset \RR^\AmbDim$ ``with tangent planes,'' that need not even be manifolds~\cite{Strzelecki:2013:JGA}.  In the case of 2-dimensional surfaces, one can argue (as discussed below) that $d\EnergyP$ is a nonlocal, nonlinear differential operator of \emph{fractional} order $2(2 - 2/p) \in ]3,4[$, rather than integer order.  This distinguishes the tangent-point energy from standard geometric energies like Willmore, and it is why we have to develop special tools for it.

In more detail: \citet{Strzelecki:2013:JGA} show that if tangent-point energy is finite for some $\DomDim$-dimensional \(\varSigma \subset \RR^\AmbDim\), then $\varSigma$ must be an \emph{embedded} submanifold of H\"{o}lder class $C^{1,\alpha}$, where $\alpha = 2 - 2 \DomDim/p$.  Intuitively: it must be free of self-intersections, and also fairly regular.  This result is improved by \cite{Blatt:2013:EST}, who establishes that $\EnergyP$($\varSigma$) is finite if and only if $\varSigma$ is an embedded submanifold of \emph{fractional Sobolev class} $W^{s,p}$, where $s = 2 - \DomDim/p$. In particular, this implies that $\varSigma$ can be expressed as an embedding $f \in W^{s,p}(M;\RR^\AmbDim)$ for some smooth manifold $M$.  For $\DomDim=2$ we have $s \in ]3/2,2[$, so we inevitably have to deal with fractional Sobolev spaces.  Knowing the natural habitat of $\EnergyP$ is key because it allows for the following observation: the differential $d\EnergyP$ is a mapping from $W^{s,p}$ to the dual space $(W^{s,p})^* = W^{-s,p}$. Hence it is plausible that $d\EnergyP$ reduces the differentiability of its argument by $2s = 2(2-2/p)$, as claimed above.

\subsection{Inner Product}
\label{sec:FractionalLaplacian}

\begin{algorithm}[t]
 \caption{Assembly of the exact discrete fractional operator $L^\sigma$}
 \label{alg:DiscreteLaplacianAssembly}
\SetAlgoLined
 initialize $L^\sigma \gets 0$ \\
 \ForAll{distinct pairs of faces $S,T$}{
  \ForAll{vertices $i$ adjacent to $S$ or $T$}{
   \ForAll{vertices $j$ adjacent to $S$ or $T$}{
	$L^\sigma_{ij} \gets L^\sigma_{ij} + 
    \frac{
    		(
    			\FaceAv{\phi}_i(S) - \FaceAv{\phi}_i(T)
    		)
		(
    			\FaceAv{\phi}_j(S) - \FaceAv{\phi}_j(T)
    		)
    	}{
    		\nabs{\Bary{f}(S) - \Bary{f}(T)}^{2\sigma + 2}
    	} \ a_f(S)\ a_f(T)$
   }
  }
 }
 \Return{$L^\sigma$}
\end{algorithm}

Standard (integer) Sobolev inner products are expressed via the Laplacian \(\Delta\).  We likewise consider the \emph{fractional Laplacian} of order $0 < 2\sigma < 2$ on $\RR^\DomDim$, which can be expressed in integral form up to a constant factor as
\begin{equation}
\label{eq:FractionalLaplacian}
\innerprod{(-\Delta)^{\sigma}u, v}_{L^2} = \iint_{\RR^\DomDim} \frac{(u(x) - u(y)) \, (v(x) - v(y))}{\nabs{x - y}^{2\sigma + \DomDim}} \ dx \ dy
\end{equation}
for sufficiently smooth functions $u,v : \RR^\DomDim \to \RR$ \cite{Kwa:2017}.
While this formula only relates to $\RR^\DomDim$, we can obtain an analogous operator $L^\sigma$ of fractional order $2 \sigma$ on functions $u, \ v : M \to \RR$ by mimicking this expression on the $\DomDim$-dimensional manifold $M$:
\begin{equation}
\label{eq:FractionalSurfaceLaplacian}
\innerprod{L^\sigma u, v}_{L^2} = \iint_{M^2} \frac{(u(x) - u(y)) \, (v(x) - v(y))}{\nabs{f(x) - f(y)}^{2\sigma + \DomDim}} \ dx_f \ dy_f .
\end{equation}
Note that, for $p > 2\DomDim$, the order of $d\EnergyP$ is $2s = 2(2-\DomDim/p) > 3$, which is outside the bounds $0 < 2\sigma < 2$.
We can ``boost'' the order of this operator by introducing a first order derivative operator $\Df$ in the numerator, yielding a ``high-order'' operator
\begin{equation}
\label{eq:HighOrderTerm}
\innerprod{Bu, v}_{L^2} \!= \!\iint_{M^2}\!\! \frac{\innerprod{\Df u(x) - \Df u(y), \Df v(x) - \Df v(y)}}{\nabs{f(x) - f(y)}^{2\sigma + \DomDim}} dx_f dy_f.
\end{equation}
More precisely, we use $\Df u(x) := d u(x) \, d f(x)^{\dagger} \in \operatorname{End}(\RR^\AmbDim)$, where $d f(x)^{\dagger} \in \operatorname{Hom}(\RR^\AmbDim; T_xM)$ denotes the Moore-Penrose pseudoinverse of $d f(x)$.
If we now let $\sigma = s - 1$, then the operator $B$ achieves the desired order $2s$.

\paragraph{Low order term.} As proposed by \citet{Yu:2021:TOG}, we can get even better preconditioning in situations with close contact by adding an additional term of lower order, which in our case translates to
\begin{equation}
\label{eq:LowOrderTerm}
\innerprod{B_0 u, v}_{L^2} \! = \!\iint_{M^2} \!\!\frac{(u(x) - u(y))(v(x) - v(y))}{\nabs{f(x) - f(y)}^{2\sigma + \DomDim}} k_{f,2}(x, y)  dx_f  dy_f.
\end{equation}
The inclusion of the tangent-point kernel $k_{f,2}(x, y)$ effectively distorts lengths in regions of high energy: as the local energy increases, so too does the apparent length induced by the inner product. As a result, self-intersecting configurations, having infinite energy, are so distant (if not infinitely so) that they are unlikely to be reached within a finite time.
The kernel $k_{f,2}(x, y)$ is chosen here so that $B$ and $B_0$ have the same units and thus behave similarly under scaling.

The overall operator $A = B + B_0$ will define the inner product we consider throughout this work. The order of this inner product matches that of the Sobolev space $W^{s,2} = H^s$, so we will occasionally use the term $H^s$ to refer to our preconditioner.

\section{Discretization}
\label{sec:Discretization}

Here, we present discretizations of all components needed for our surface optimization scheme.  The basic idea is to minimize tangent-point energy by following the gradient flow, preconditioned by our fractional inner product.  In practice we will also want to incorporate a variety of constraints, which we do by both projecting the flow direction onto the tangent space of the constraint manifold, and by then projecting the surface itself onto this manifold.  The overall algorithm for each descent step can be summarized as:
\begin{enumerate}
\item Assemble the derivative $\DDiscEnergyP(f)$ of the energy (\secref{DiscreteTangentPointEnergy}).
\item Construct the fractional operator $A = B + B_0$ (\secref{DiscreteFractionalLaplacian}).
\item Solve \eqref{SaddleProblem} to obtain the descent direction $\mathbf{x}$.
\item Take a step in the direction of $\mathbf{x}$ using Armijo line search.
\item Project the resulting embedding onto the constraint manifold of $\Phi$ (\secref{CorrectiveProjection}).
\end{enumerate}
As noted in \secref{Introduction}, the initial algorithm outlined in this section is quite inefficient; we will introduce accelerations in subsequent sections. For the final accelerated algorithm, see \secref{AcceleratedAlgorithmOverview}.

\subsection{Discrete Energy}
\label{sec:DiscreteTangentPointEnergy}

On a discrete triangle mesh $M = (V, E, F)$ with embedding $f : V \to \RR^3$, we evaluate the double integral of \eqref{TangentPointEnergy} using simple mid-point quadrature on all faces. We define the \emph{discrete tangent-point kernel} on a pair of faces $S, T$ as
\begin{equation}
	\label{eq:DiscreteTangentPointKernel}
	K_{f,p}(S, T) 
	= 
	\frac{\nabs{\Proj_f(S) \, (\Bary{f}(S) - \Bary{f}(T))}^{p}}{\nabs{\Bary{f}(S) - \Bary{f}(T)}^{2p}},
\end{equation}
where $\Bary{f}(S)$ denotes the barycenter of face $S$ under embedding $f$.
The full energy is then defined as a double sum over faces
\begin{equation}
	\DiscEnergyP(f) = \sum_{S \in F} \sum_{T \in F} K_{f,p}(S, T)\ a_f(S)\ a_f(T),
	\label{eq:DiscreteEnergy}
\end{equation}
where $a_f(S)$ denotes the area of face $S$ under embedding $f$. The differential $\DDiscEnergyP(f)$ of this energy with respect to $f \in \RR^{3\nabs{V}}$ can be obtained via the chain rule.

\subsection{Discrete Inner Product}
\label{sec:DiscreteFractionalLaplacian}

The fractional operator $L^\sigma$ can be discretized as a $\abs{V} \times \abs{V}$ matrix with entries obtained from the right-hand side of \eqref{FractionalSurfaceLaplacian}. 
The rows and columns of $L^{\sigma}$ are indexed by vertices, and each entry can na\"{i}vely be computed as
\begin{equation}
\label{eq:DiscreteLaplacianQuartic}
	L^{\sigma}_{ij} 
	= 
	\mathop{\sum_{S \in F} \sum_{T \in F}}_{S \neq T} 
	\frac{
		(
			\FaceAv{\phi}_i(S) - \FaceAv{\phi}_i(T)
		)
		\,
		(
			\FaceAv{\phi}_j(S) - \FaceAv{\phi}_j(T)
		)
	}{
		\nabs{\Bary{f}(S) - \Bary{f}(T)}^{2\sigma + 2}
	}\ a_f(S)\ a_f(T) 
	,
\end{equation}
where
$\phi_i$ denotes the piecewise linear hat-function centered at vertex $i$, and where 
$\FaceAv{\phi}_i(S)$ denotes its evaluation on the barycenter of $S$, \ie, $\FaceAv{\phi}_i(S)$ is $1/3$ if vertex $i$ is adjacent to face $S$ and $0$ otherwise.
Assembling $L^{\sigma}$ using \eqref{DiscreteLaplacianQuartic} would require quartic complexity; however, the integrand vanishes for most pairs $S,T$, so the assembly can be done in quadratic time by only considering nonzero contributions (\algref{DiscreteLaplacianAssembly}).

\subsubsection{High- and Low-Order Terms}
\label{sec:HighAndLowOrderTerms}

The high-order matrix $B$ of our inner product (\eqref{HighOrderTerm}) can be assembled using the same procedure as in \algref{DiscreteLaplacianAssembly}, simply using the summand
\begin{equation}
\label{eq:DiscreteHighOrderTerm}
\frac{\innerprod{\DMatrix \phi_i(S) - \DMatrix \phi_i(T), \DMatrix \phi_j(S) - \DMatrix \phi_j(T)}}{\nabs{\Bary{f}(S) - \Bary{f}(T)}^{2\sigma + 2}}\ a_f(S)\ a_f(T)
\end{equation}
in place of the one in \eqref{DiscreteLaplacianQuartic}. Here $\DMatrix$ is a discretization of $\Df$. Intuitively, $\DMatrix u(S)$ is the derivative of the function $u = \sum_{i\in V} u_i \, \phi_i$ within the triangle $S$, and can be evaluated as
\begin{align*}
	- 2 \, a_f(S)^{-1}
\big(
	N_f(S) \times (u_i e_{jk} + u_j e_{ki} + u_k e_{ij})
\big)^\T
	,
\end{align*}
where $i,j,k$ are the vertices of triangle $S$ and where $e_{jk}$, $e_{ki}$, and $e_{ij})$ denote the unit edge vector of $S$.
The low-order matrix $B_0$ can be assembled likewise with the summand
\begin{equation}
\label{eq:DiscreteLowOrderTerm}
	\frac{
		(
			\FaceAv{\phi}_i(S) - \FaceAv{\phi}_i(T)
		)
		\,
		(
			\FaceAv{\phi}_j(S) - \FaceAv{\phi}_j(T)
		)
	}{
		\nabs{\Bary{f}(S) - \Bary{f}(T)}^{2\sigma + 2}
	}
	\ K_{f,2}(S, T) \ a_f(S)\ a_f(T)
\end{equation}
using the discrete tangent-point kernel $K_{f,2}(S, T)$ from \eqref{DiscreteTangentPointKernel}. We can then assemble $A = B + B_0$ by assembling both terms. The matrix $A$ is $\abs{V} \times \abs{V}$ and thus applies to scalar functions, but we can construct a corresponding operator $A_3$ on vector-valued functions $u, v : V \to \RR^3$ by replacing each entry $A_{ij}$ with the $3 \times 3$ block $A_{ij} \, \mathsf{I}_{3 \times 3}$, thus obtaining a matrix of size $3\abs{V} \times 3\abs{V}$.

\subsection{Constraints}
\label{sec:ConstrainedGradientProjection}

\subsubsection{Gradient Projection}
Like the integer vector Laplacian, our operator $A_3$ possesses a nullspace consisting of uniform translations.
A simple way to eliminate this nullspace is to define a constraint function $\ConstraintFn : \RR^{3\abs{V}} \to \RR^\ConstraintDim$ and require (with some abuse of notation) that $\ConstraintFn(f) = 0$. The constrained descent direction $\mathbf{x}$ can then be obtained by solving the saddle point problem
\begin{equation}
\label{eq:SaddleProblem}
\left[
\begin{array}{cc}
A_3 & \DConstraintFn(f)^{\T} \\
\DConstraintFn(f) & 0
\end{array}
\right]
\left[
\begin{array}{c}
\mathbf{x} \\
\lambda
\end{array}
\right] = \left[
\begin{array}{c}
-\DDiscEnergyP(f) \\
0
\end{array}
\right]
	,
\end{equation}
where $\DConstraintFn(f)$ denotes the Jacobian of $\ConstraintFn$.
Assuming a suitable $\ConstraintFn$ is chosen, this saddle point matrix is invertible, and the result will be tangent to the constraint manifold $\{ f \mid \ConstraintFn(f) = 0 \}$. Beyond just eliminating nullspaces, such constraints can also be used to achieve design objectives such as control of areas or volumes.

\subsubsection{Corrective Projection}
\label{sec:CorrectiveProjection}

The descent direction $\mathbf{x}$ obtained in \secref{ConstrainedGradientProjection} is tangent to the constraint manifold, but this does not prevent the embedding \(f\) itself from drifting away from the constraint manifold.  To counteract this, after we have found a feasible step size $\tau >0$ via line search, we project the current state $f + \tau \, \mathbf{x}$ back onto the constraint manifold of $\ConstraintFn$.
We reuse the left-hand side of \eqref{SaddleProblem} and solve
\begin{equation}
\label{eq:ProjectionSaddleProblem}
\left[
\begin{array}{cc}
A_3 & \DConstraintFn(f)^{\T} \\
\DConstraintFn(f) & 0
\end{array}
\right]
\left[
\begin{array}{c}
\mathbf{h} \\
\lambda
\end{array}
\right] = \left[
\begin{array}{c}
0 \\
-\ConstraintFn(f + \tau \, \mathbf{x})
\end{array}
\right]
\end{equation}
to obtain a Newton step $\mathbf{h}$, which we add to the updated embedding $f + \tau \, \mathbf{x}$.
With respect to the metric encoded by $A_3$, $\mathbf{h}$ is the least-norm solution of the linear equation $\DConstraintFn(f)\, \mathbf{h} = -\ConstraintFn(f)$.
This correction can be repeated several times if the constraint violation is not sufficiently close to $0$.
For the constraints we explored, however, a single step was always sufficient.

\section{Fast Energy and Derivative Evaluation}
\label{sec:BarnesHut}

The na\"{i}ve algorithm of \secref{Discretization} is bottlenecked by several operations of at least quadratic complexity. The first such bottleneck is the evaluation of the energy and its derivative, which requires iteration over all pairs of elements.
We thus use a Barnes-Hut hierarchical approximation~\cite{BarnesHFCA1986} to evaluate the tangent-point energy $\DiscEnergyP$ and its derivative $\DDiscEnergyP$.

\subsection{Approximate Energy}
\label{sec:ApproximateEnergy}

The kernel $K_{f,p}(S,T)$ (\eqref{DiscreteTangentPointKernel}) only requires three quantities to evaluate: the barycenters of $S$ and $T$, and the normal projector of $S$. We can make this dependence clearer by rewriting it as 
$K_{f,p}(S,T) = K_{p}(\Bary{f}(S), \Proj_f(S); \Bary{f}(T))$, with
\begin{equation*}
K_{p}(X, P; Y) := \frac{\nabs{P \, (X - Y)}^{p}}{\nabs{X - Y}^{2p}}.
\end{equation*}
We can then hierarchically approximate the all-pairs interactions of $K_p$.
We construct a bounding-volume hierarchy (BVH) on the face set $F$, where each node $\Ical$ computes the total area $a_{\Ical}$ and the barycenter $X_{\Ical}$ of its elements. To reduce the number of nodes, we stop splitting leaf nodes once they have $l$ or fewer elements ($l=8$ in our experiments).
For a given $\theta \geq 0$, we say that $\Ical$ is \emph{admissible} with respect to $S$ if (1) it is a leaf node or if (2) it satisfies 
\begin{align*}
	\max(r(S), r(\Ical)) < \operatorname{dist}( S, \operatorname{conv}(\Ical)).
\end{align*}
Here $r(S)$, $r(\Ical)$ are the radii of the triangle $S$ and the node $\Ical$, respectively, both measured from their barycenters;
$\operatorname{dist}$ denotes the minimal Euclidean distance between two sets; and
$\operatorname{conv}(\Ical)$ denotes the convex hull of $\Ical$.
In practice, we approximate these quantities by replacing the node $\Ical$ by its axis-aligned bounding boxes, leading to a slightly stricter admissability condition.
Then, $\text{adm}(S)$ is the set of all admissible nodes with respect to $S$ with no admissible ancestors. 
The energy evaluation then becomes the sum
\begin{equation}
	\ApproxDiscEnergyP(f) 
	= 
	\sum_{S \in F} \sum_{\Ical \in \text{adm}(S)} 
	K_p(\Bary{f}(S), \Proj_f(S); X_\Ical)\ a_f(S)\ a_{\Ical}.
	\label{eq:BarnesHutEnergy}
\end{equation}
 The separation parameter $\theta$ controls the approximation quality; the higher $\theta$ is, the faster the computation, but the less accurate the result. For $\theta = 0$, the sum degenerates to an all-pairs exact computation. Unless otherwise noted, we use $\theta = 0.5$ for all experiments.

\subsection{Approximate Derivative}
\label{ApproximateGradient}

Computing an approximate derivative with Barnes-Hut is not entirely analogous to computing the energy. For each vertex $v \in V$, we evaluate the sum
\begin{equation*}
	\widetilde{\partial_v} \DiscEnergyP(f) 
	= 
	\sum_{S \in F(v)} \sum_{\Ical \in \text{adm}(S)} 
	\partial_v
\big(K_p(\Bary{f}(S), \Proj_f(S); X_{\Ical})\ a_f(S)\ a_{\Ical} \big)
	,
\end{equation*}
where $F(v)$ denotes the set of faces containing $v$.
This approximates both the forward and reverse terms that would be differentiated by $v$ in an exact computation. Note that the outer sum over all $S \in F(v)$ for both energy and derivative evaluations can be evaluated as a parallel reduction without modification.

\section{Iterative Linear Solver}
\label{sec:PreconditionedIterativeMethod}
An even more significant bottleneck than the energy is the dense saddle point problem of \eqref{SaddleProblem}. 
Rather than solving this problem via dense matrix inversion, we will solve it instead using GMRES, an iterative method. In general, efficient iterative methods require two key ingredients: fast matrix-vector products, and effective preconditioners. Here, we will describe methods for both.

\subsection{Hierarchical Matrices}
\label{sec:HierarchicalMatrices}

We use hierarchical matrices~\cite{Hackbusch:2015:HMA} to perform fast multiplication with $A$ without explicitly assembling the matrix. In this section, we present the special case of rank-1 compression of kernel matrices, while noting that the original method can also perform higher-rank approximations. 
In our setting, a \emph{kernel matrix} $\KernelMatrix$ is a matrix of size $\nabs{F} \times \nabs{F}$ whose entries are defined by
\[
	\KernelMatrix_{ST} 
	= 
	(1 - \delta_{ST}) \, h(\Bary{f}(S), \Proj_f(S); \Bary{f}(T) , \Proj_f(S)),
\]
where $h : (\RR^\AmbDim  \times \operatorname{End}(\RR^\AmbDim))  \times (\RR^\AmbDim \times \operatorname{End}(\RR^\AmbDim)) \to \RR$ is a suitable kernel function.
To motivate this approach, we first reduce the actions of the operators $L^\sigma$, $B$, and $B_0$ to the multiplication with certain kernel matrices.

\subsubsection{Applying the operator $L^\sigma$}
\label{sec:ApplyingTheFractionalLaplacian}
An elementary computation shows (see \appref{ActionOfTheFractionalLaplacian}) that the action of the discrete linear operator $L^{\sigma}$ on a vector $\mathbf{v} \in \RR^{\nabs{V}}$ 
can be written as
\[
	L^{\sigma} \mathbf{v} 
	= 
	2\, 
	\PreLo^{\T}
	\big[\diag \big( \diag(a_f)^{-1} \KernelMatrix \, a_f \big) - \KernelMatrix \big] 
	\, 
	\PreLo
	\,
	\mathbf{v}
	.
\]
Here $a_f$ is the $\nabs{F}$-vector of face areas; 
$\PreLo$ is the $\nabs{F} \times \nabs{V}$-matrix that averages values on vertices onto faces and multiplies with the face areas;
and $\KernelMatrix$ is the kernel matrix of size $\nabs{F} \times \nabs{F}$ to the singular kernel $\Kernel (X, P; Y, Q) = \nabs{X-Y}^{-(2\sigma + 2)}$.
$\PreLo$ is sparse, so we just need an efficient product with $\KernelMatrix$ to evaluate the full product with $L^\sigma$.

\subsubsection{Applying the High-Order Term}
\label{sec:ApplyingTheHighOrderTerm}
To evaluate a matrix-vector product with $A = B + B_0$, it suffices to evaluate $B$ and $B_0$ separately.
This can be done in a similar fashion as for $L^\sigma$.
For the higher order term $B$, we have the identity
\[
	B \mathbf{v} 
	= 
	2\, 
	\PreHi^{\T}
	\big[\diag \big( \diag(a_f)^{-1} \KernelMatrix \, a_f \big) - \KernelMatrix \big] 
	\, 
	\PreHi
	\,
	\mathbf{v}
	,
\]
where $\PreHi = \diag(a_f) \, \DMatrix$
with the discrete derivative operator $\DMatrix$ described in \secref{HighAndLowOrderTerms}
and where the kernel $\Kernel$ of the kernel matrix $\KernelMatrix$ is given by
$h(X,P; Y,Q) =  \nabs{X-Y}^{-(2(s-1) + 2)}$.

\subsubsection{Applying the Low-Order Term}
\label{sec:ApplyingTheLowOrderTerm}
Likewise, we can write the action of $B_0$ as
\[
	B_0 \mathbf{v} 
	=
	2\, 
	\PreHi^{\T}
	\big[\diag \big( \diag(a_f)^{-1} \KernelMatrix \, a_f \big) - \KernelMatrix \big] 
	\, 
	\PreHi
	\,
	\mathbf{v}
	,
\]
where the kernel $\Kernel$ of the kernel matrix $\KernelMatrix$  is given by
\[
	\Kernel(X,P; Y,Q) 
	=  
	\frac{k_{2}(X, P; Y) + k_{2}(Y, Q; X) }{2 \nabs{X-Y}^{2(s-1) + 2}}
	.
\]

\subsubsection{Block Cluster Tree}
\label{sec:BlockClusterTree}

In order to compress these kernel matrices, we reuse the BVH from \secref{BarnesHut}, but additionally compute the average projector $\Proj_\Ical := a_\Ical^{-1} \sum_{S \in \Ical} a_f(S) \, P_f(S)$ for each node $\Ical$.
From this, we construct a \emph{block cluster tree}, whose nodes (termed \emph{block clusters}) consist of pairs of BVH nodes (termed \emph{clusters in the following}).
For a given separation parameter $\chi \geq 0$, we say that two BVH clusters $\Ical$ and $\Jcal$ are an \emph{separated pair} if 
\begin{align*}
	\max\left(r(\Ical), r(\Jcal)\right)  \leq \chi \,  \mathrm{dist}( \mathrm{conv}(\Ical),\mathrm{conv}(\Jcal) ).
\end{align*}
\setlength{\columnsep}{1em}
\setlength{\intextsep}{0em}
\begin{wrapfigure}{r}{110pt}
   \includegraphics[width=110pt]{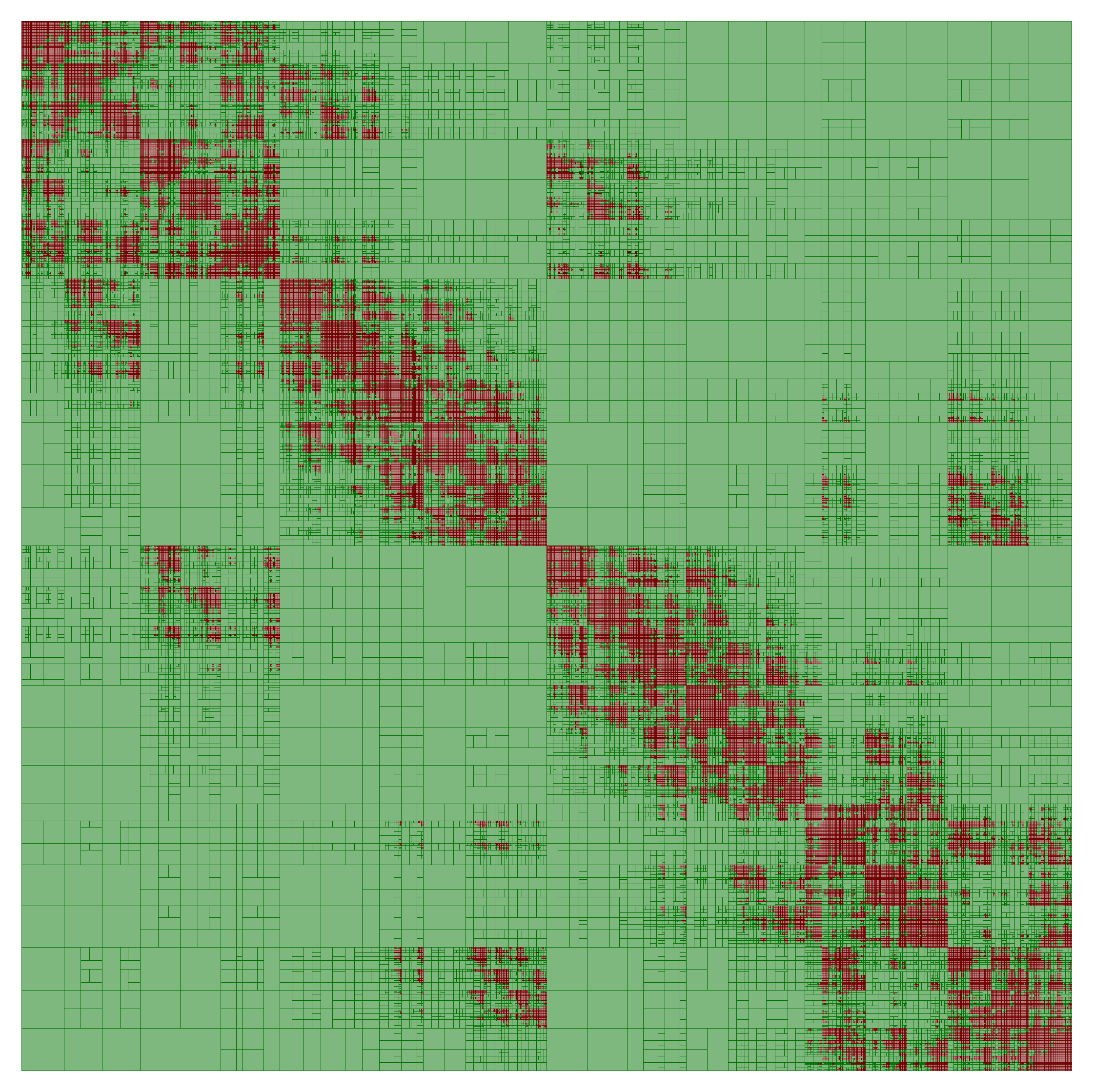}
   \vspace{-2\baselineskip}\caption{A block decomposition of a kernel matrix $\KernelMatrix$ induced by a block cluster tree. Admissible blocks are shown in green, while inadmissible blocks are in red.\label{fig:BCT}}
\end{wrapfigure}
Here again, $r(\Ical)$, $r(\Jcal)$ are the radii of the nodes $\Ical$, $\Jcal$ as measured from their barycenters. The parameter $\chi$ controls the accuracy of the approximation; it will be discussed further in the next section.
Then, denoting the BVH root by $\mathcal{R}$, we construct the block cluster tree by starting with the single pair $(\mathcal{R}, \mathcal{R})$, and iteratively splitting nonseparated nodes $(\Ical, \Jcal)$ into the Cartesian products of their constituents' children until all leaf nodes are either separated or cannot be split any further.
In practice, the tree structure is not important to maintain; only the lists of leaf nodes matter.
We refer to the separated leaf nodes of the block cluster tree as \emph{admissible blocks} and to the others as \emph{inadmissible blocks}; \figref{BCT} illustrates the decomposition of the full matrix into these blocks.

\subsubsection{Hierarchical Multiplication}
\label{sec:HierarchicalMultiplication}

The block cluster tree allows us to perform approximate multiplication with a \emph{kernel matrix} $\KernelMatrix$ as follows.
Every pair of BVH clusters $(\Ical, \Jcal)$ corresponds to a block of $\KernelMatrix$ with rows indexed by $\Ical$ and columns by $\Jcal$. 
Let $\KernelMatrix_{\Ical\Jcal}$ denote this matrix block and let
$\mathbf{x}_\Ical$ and  $\mathbf{1}_\Ical$ denote the slices of $\mathbf{x} \in \RR^{\nabs{F}} $ and of the all-ones vector indexed by $\Ical$, respectively. 
Then, for all leaf blocks $(\Ical, \Jcal)$, we compute the product $\mathbf{y} = \KernelMatrix \, \mathbf{x}$ in two steps:
\begin{enumerate}
\item If $( \Ical, \Jcal )$ is inadmissible, then we multiply exactly:
\[
	\mathbf{y}_\Ical
	\gets 
	\mathbf{y}_\Ical
	+ 
	\KernelMatrix_{\Ical\Jcal} \, \mathbf{x}_\Jcal
	.
\]
\item If $( \Ical, \Jcal )$ is admissible, we employ rank-one approximation:
\[
	\mathbf{y}_\Ical
	\gets 
	\mathbf{y}_\Ical
	+ 
	\mathbf{1}_\Ical \,  
	\Kernel
		(X_{\Ical}, \Proj_{\Ical}  ;  
		X_{\Jcal}, \Proj_{\Jcal}
	) 
	\, 
	\mathbf{1}_\Jcal^{\T} \, \mathbf{x}_\Jcal
	.
\]
\end{enumerate}
Here, we can see more clearly the effect of $\chi$. For $\chi = 0$, all blocks are considered inadmissible, and the action of $\KernelMatrix$ is evaluated exactly. For $\chi > 0$, the larger the value, the more blocks will be considered admissible and thus multiplied using the fast approximation in Step~2, leading to faster evaluation time -- but also higher error, analogous to the $\theta$ parameter for Barnes-Hut. For our experiments, we found $\chi = 0.5$ to be a broadly acceptable value. Note that, while a straightforward implementation of these two steps is sufficient to evaluate the product, a much faster implementation can be obtained by employing multipole methods; see \appref{FastMatrixVectorMultiplication} for details.

\subsection{Preconditioner}
\label{sec:Preconditioner}

While we can now evaluate matrix-vector products with $A$ efficiently, this alone does not generally allow us to efficiently solve $Ax = b$. We further require a \emph{preconditioner} whose action can be computed efficiently. As we never construct $A$, classical preconditioners such as incomplete Cholesky factorizations or even $\diag(A)$ are unusuable.
Instead, we note that our operator $A$ is closely related to the fractional Laplacian $(-\Delta_M)^s$, and has the same order $2s$. Assembling $(-\Delta_M)^s$ is infeasible, but we can obtain a cheap approximation of its inverse $(-\Delta_M)^{-s}$ by factoring it as
\begin{equation*}
	(-\Delta_M)^{-s} 
	= 
	(-\Delta_M)^{-1} (-\Delta_M)^{2-s} (-\Delta_M)^{-1}
	,
\end{equation*}
where the two occurrences of the integer Laplace-Beltrami operator $(-\Delta_M)$ can then be replaced by the sparse cotan-weighted Laplace-Beltrami operator on meshes. 
What remains is a forward application of the fractional Laplacian $(-\Delta_M)^{2-s}$, to which we do not have direct access. Fortunately, since $0 < 2-s < 2$ holds, we can replace $(-\Delta_M)^{2-s}$ with $L^{2-s}$ (as per \secref{FractionalLaplacian}), whose action can efficiently approximated by \secref{ApplyingTheFractionalLaplacian}.
Thus, if we first pre-factorize $(-\Delta_M)$, we can apply our preconditioner
\begin{equation*}
	\widetilde{A}^{-1} := (-\Delta_M)^{-1} L^{2-s} (-\Delta_M)^{-1}
\end{equation*}
with just two back-substitutions and one hierarchical matrix-vector product per application, all of which can be evaluated reasonably efficiently. Note that, despite having the same order as our operator $A$ (and therefore our energy), $\widetilde{A}^{-1}$ is not suitable for direct use as the inner product: as a direct approximation of the inverse operator (as opposed to the forward operator), it cannot be added with other inner product terms such as those of \eqref{LowOrderTerm} or \secref{WillmoreEnergy}. As a preconditioner for GMRES, however, it is highly effective, allowing us to invert $A$ (plus any auxiliary terms) efficiently.

\subsection{Schur Complement}
\label{sec:SchurComplement}

While we are now capable of solving the \emph{unconstrained} problem $Ax = b$ iteratively, this does not immediately allow us to solve the saddle point problem (\secref{ConstrainedGradientProjection}). While the method can be applied, we empirically found that it exhibited poor convergence when used on the constrained system. We instead use the \emph{Schur complement}~\cite{Zhang05} to handle the additional rows. Let $M$ be the saddle point matrix:
\[
M :=
\left[
\begin{array}{cc}
A_3 & \DConstraintFn^{\T} \\
\DConstraintFn & 0
\end{array}
\right]
\]
Then, the Schur complement of $M$ with respect to $A_3$ is given by
\begin{equation}
\label{eq:SchurComplementMA}
(M / A) = -\DConstraintFn \, \big(A_3^{-1} \DConstraintFn^{\T}\big)
\end{equation}
Note that it is useful to cache $A_3^{-1} \DConstraintFn^{\T}$ here for future reuse. Expressions for each block of $M^{-1}$ are then given as
\[
\left[
\begin{array}{cc}
	A_3^{-1} + \big( A_3^{-1} \DConstraintFn^{\T} \big) (M/A_3)^{-1} \DConstraintFn \, A_3^{-1}
	& -\big(A_3^{-1}  \DConstraintFn^{\T}\big) (M/A)^{-1} 
	\\
	-(M/A)^{-1} \big( A_3^{-1} \DConstraintFn^{\T} \big)^\T
	& (M / A)^{-1}
\end{array}
\right]
\]
$A_3^{-1}$ can be applied using the iterative method just outlined; a product with $A_3$ is equivalent to three separate products with $A$.
The complement $M/A$ is dense, but it has dimensions $\ConstraintDim \times \ConstraintDim$, corresponding to the number of scalar constraints. As long as $\ConstraintDim$ is a small constant, $(M/A)^{-1}$ can be computed quickly.
Thus, all blocks of $M^{-1}$ can be computed without having to invert a large matrix.
Further, to obtain the constrained descent direction $\mathbf{x}$, we only require the top-left block. Let $\mathbf{g} := \DDiscEnergyP(f)$; then, we can compute the descent direction by directly applying the top-left block to $\mathbf{x}$, producing
\begin{equation}
\label{eq:SchurProjectedExpression}
\mathbf{x} = A_3^{-1} \mathbf{g} + \big(A_3^{-1} \DConstraintFn^{\T}\big) (M/A)^{-1} \DConstraintFn \, \big( A_3^{-1} \mathbf{g} \big).
\end{equation}
\eqref{SchurComplementMA} requires one application $A_3^{-1}$ per row of $d\Phi$. \eqref{SchurProjectedExpression} contains three occurrences of $A_3^{-1}$, but $A_3^{-1} \mathbf{g}$ can be reused in both places where it appears, and $A_3^{-1} d\Phi^{\T}$ can be reused from its earlier computation in \eqref{SchurComplementMA}. Thus, the method requires $\ConstraintDim + 1$ iterative solves, where $\ConstraintDim$ is the number of constraints. In our examples, we never have $\ConstraintDim > 2$, so the cost remains acceptable.

\subsubsection{Corrective Projection}
\label{sec:SchurCorrectiveProjection}

We similarly use the Schur complement to solve \eqref{ProjectionSaddleProblem} for the corrective step $\mathbf{h}$. Only the top-right block of the Schur complement is needed, giving the expression
\begin{equation}
\mathbf{h} = - \big(A_3^{-1} \DConstraintFn^{\T}\big) (M/A)^{-1} (-\ConstraintFn(f)).
\end{equation}
$(M/A)$ does not need to be recomputed, and $A_3^{-1} \DConstraintFn^{\T}$ can again be reused. Thus, constraint projection incurs no significant costs.

\subsection{Accelerated Algorithm Overview}
\label{sec:AcceleratedAlgorithmOverview}

The accelerated algorithm is as follows:

\begin{enumerate}
\item Assemble the (approximate) derivative $\DDiscEnergyP(f)$ of the energy using Barnes-Hut (\secref{BarnesHut}).
\item Construct a BVH that partitions the faces of the mesh, and use it to create a block cluster tree (\secref{BlockClusterTree}).
\item Use the Schur complement to solve the constrained saddle point problem (\eqref{SaddleProblem}).
\begin{enumerate}
  \item Evaluate products with $(A_3)^{-1}$ by using a matrix-free iterative method (\eg{}, GMRES), with the preconditioner from \secref{Preconditioner}, and an initial guess of 0.
  \item Within the iterative method, evaluate products with $A$ using the block cluster tree (\secref{ApplyingTheHighOrderTerm}, \secref{ApplyingTheLowOrderTerm}).
\end{enumerate}
\item Take a step in the direction of $\mathbf{x}$ using standard line search.
\item Reuse the Schur complement to project the resulting embedding onto the constraint manifold of $\Phi$ (\secref{SchurCorrectiveProjection}).
\end{enumerate}
If no constraints are imposed, then the algorithm can be simplified: step 3 can be replaced by a single iterative solve $A_3 x = b$, and step 5 can be omitted entirely.

\section{Dynamic Remeshing}
\label{sec:DynamicRemeshing}

\begin{figure}
   \includegraphics[width=\columnwidth]{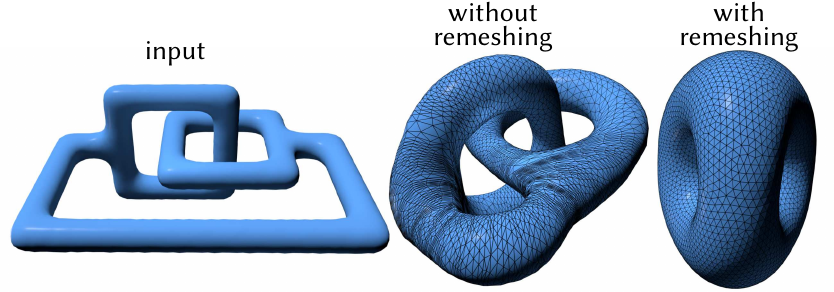}
   \caption{Adaptive remeshing not only improves element quality---it also helps to avoid local minima where the surface gets ``stuck.''\label{fig:RemeshingComparison}}
\end{figure}

Minimizing the tangent-point energy often induces large surface deformations that degrade triangle inequality. We therefore use a dynamic remeshing scheme similar to the approach of~\citet{Chen:2011}. The exact algorithm we use is as follows:
\begin{enumerate}
\item Edges with length greater than $3L_0 / 2$ are split and edges with length smaller than $L_0 / 2$ are collapsed, unless this operation would result in triangle foldover.
\item For $N$ iterations: 
\begin{enumerate}
	\item All edges that violate the Delaunay condition are flipped until no such flippable edges can be found.
	\item Vertex positions are smoothed by computing a displacement vector from neighboring triangles 
\begin{align*}
	u_i = 
	\rho \,
	\frac{
		\sum_{S \in F(i)} a_f(S) (c_f(S) - f(i))
	}{
		\sum_{S \in F(i)} a_f(S)
	}.
\end{align*}
Here $F(i)$ denotes the set of faces containing vertex $i$, $c_{f}(S)$ is the circumcenter of the triangle $S$, and $\rho < 1$ is a constant. This displacement is projected onto the tangent space of the vertex and added to the original position.
\end{enumerate}
\end{enumerate}
Our implementation uses $\rho = 0.5$ and $N = 5$; $L_0$ is set to the average edge length of the initial mesh and remains constant throughout. We apply this remeshing procedure at the end of each iteration, after the final step of \secref{AcceleratedAlgorithmOverview}. Remeshing is crucial to reaching minimizers of the tangent-point energy; without it, degrading triangle quality can impede or even halt progress, as seen in \figref{RemeshingComparison}.

\section{Constraints and Penalties}
\label{sec:ConstraintsandPenalties}

A variety of constraints and penalties can be imposed on the tangent-point energy, both for regularization of minimizers and for specific design purposes. In this section, we discuss the constraints and penalties that we have investigated; more are certainly possible, and in particular, combining the tangent-point energy with other classical surface energies could make for interesting future work.

\begin{figure}[b]
   \vspace{-\baselineskip}\includegraphics[width=.85\columnwidth]{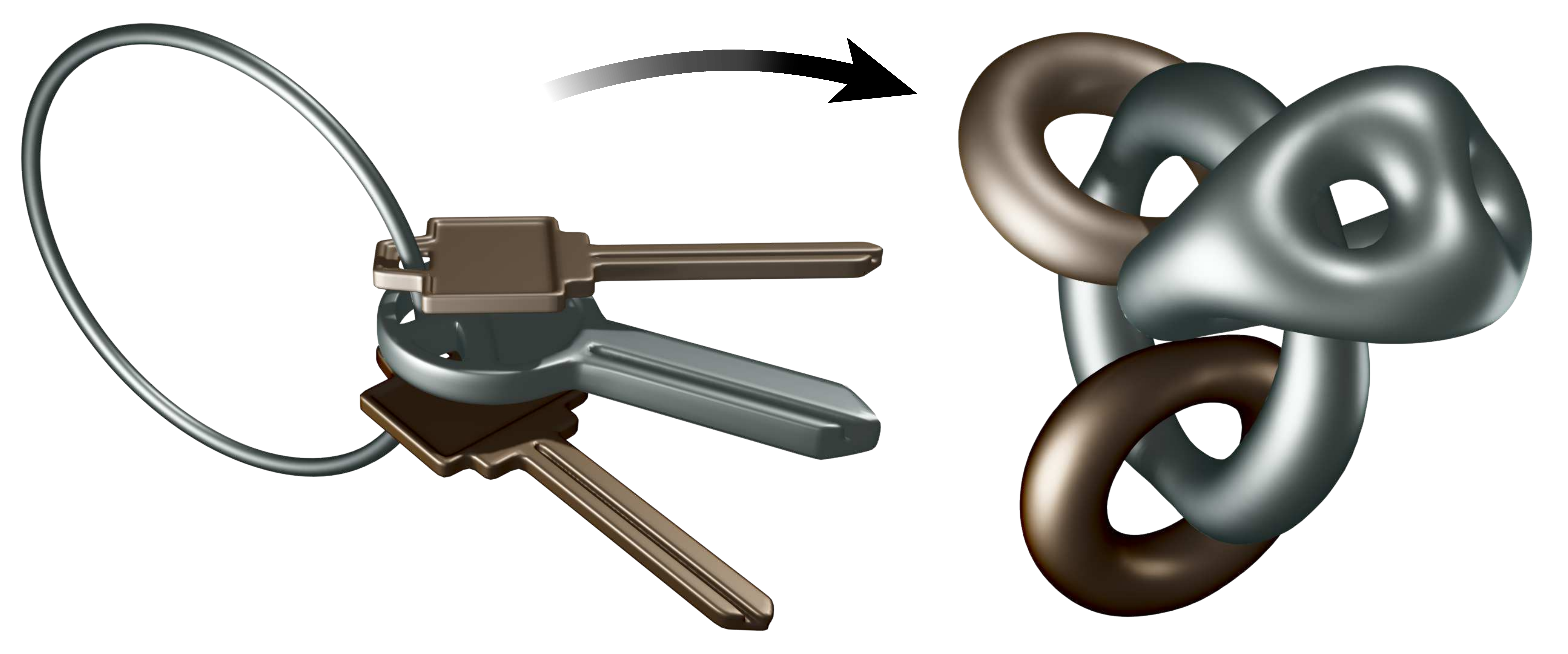}
   \caption{To handle multiple components (as shown here), we fix the barycenter of each one during preconditioning, then add back in the mean motion of each component from the original $L^2$-gradient after the solve.\label{fig:Keys}}
\end{figure}

\subsection{Constraints}

We consider four types of constraints: fixed barycenter, vertex pins, total area, and total volume.

\subsubsection{Fixed Barycenter Constraint}
\label{sec:FixedBarycenterConstraint}

A fixed barycenter constraint can be defined as 
\begin{align*}
	\Phi_{C}(f) = \frac{\sum_{i \in V} f(i) \, a_f(i)}{\sum_{i \in V} a_f(i)} - X_0, 
\end{align*}
where $X_0$ is the target barycenter location and $a_f(i)$ denotes the area associated to vertex $i$. Its Jacobian $d\Phi_C$ is a $3 \times 3|V|$ matrix consisting of $|V|$ copies of the $3 \times 3$ identity matrix appended horizontally. This constraint primarily serves to eliminate the nullspace of the fractional Laplacian (\secref{ConstrainedGradientProjection}); either a barycenter constraint or at least one pin constraint must be added to every problem to be well-posed.
For domains with multiple components, barycenters are constrained separately for each component.

\paragraph{Barycenter Motions.} In some cases, it might be desirable to allow the barycenter to float freely, \eg, when a scene contains fixed obstacles for the surface to avoid. A simple modification enables this motion: compute the weighted average over all vertices of the $L^2$ gradient before projection, and then add the constant translation by that vector back to the descent direction after projection. For domains with multiple components, the average motion is computed separately for each component (\figref{Keys}).

\subsubsection{Vertex Pin Constraints}
\label{sec:VertexPinConstraints}

A vertex pin constraint simply fixes a vertex to a position. Every pinned vertex $i$ produces a constraint function $\Phi_{P_i}(f) = f(i) - f_0(i)$, where $f_0(i)$ is the pinned position. The Jacobian $d\Phi_{P_i}$ is a $3 \times 3|V|$ matrix, but the only nonzero entries consist of a single copy of the identity matrix in the block indexed by $i$. A pin also eliminates the nullspace of the Laplacian, so if any pins are used, then a barycenter constraint is unneeded.

\subsubsection{Total Area Constraint}
\label{sec:TotalAreaConstraint}

A total area constraint preserves the total surface area of the mesh, and can be written as 
\begin{align*}
 	\Phi_A(f) = \textstyle (\sum_{T \in F} a_f(T)) - A_0,
\end{align*} 
where $A_0$ is the target area. The Jacobian $d\Phi_{A}$ is a $3|V|$ row vector with the area gradient at each vertex, which is equivalent to twice the mean curvature normal.

\subsubsection{Total Volume Constraint}
\label{sec:TotalVolumeConstraint}

Likewise, a total (signed) volume constraint can be written as 
\begin{align*}
 	\Phi_A(f) = \textstyle  \tfrac{1}{6} \big(\sum_{(ijk) \in F} \, f(i) \cdot (f(j) \times f(k)) \big) - V_0,
\end{align*} 
where $V_0$ is the target volume. For each vertex, the Jacobian $d\Phi_{A}$ is proportional to the area-weighted vertex normal.

\subsection{Fast Positional Constraints}
\label{sec:FastPositionalConstraints}

As previously discussed, computing the Schur complement requires one iterative solve per row of the constraint block $d\Phi$. For linear positional constraints such as barycenters (3 rows per component) and vertex pins (3 rows per pinned vertex), this can be disproportionately expensive. Rather than handling these rows using the Schur complement, we include them directly in the matrix $A$, producing a smaller saddle point matrix with structure analogous to \eqref{SaddleProblem}. Forward matrix-vector products for the iterative solve require only sparse products with $d\Phi_C$ and $d\Phi_{P_i}$ in addition to the hierarchical products of \secref{HierarchicalMatrices}. The same rows and columns are then appended to the integer Laplacians in the preconditioner (\secref{Preconditioner}), and the system is solved iteratively as before.

Fast convergence in this scenario requires that orthogonality to these constraints be sufficiently similar under the two inner products defined by the integer Laplacian $\Delta$ and the fractional operator $A$. Empirically, this is the case for linear positional constraints, but is \emph{not} the case for constraints such as total area and volume. Thus, we reserve the Schur complement for these more difficult constraints.

\subsection{Penalties}
\label{sec:PenaltyPotentials}

In addition to hard constraints, a number of soft penalty potentials can be added to regularize the flow in some way. These potentials are added directly to the objective function with some weighting coefficient alongside the tangent-point energy, and their gradients are accumulated in the same step.

\subsubsection{Total Area and Volume Potentials}
\label{sec:TotalAreaAndVolumePotentials}

Soft penalties for total area and volume can be used in place of hard constraints, encouraging these quantities to stay close to their initial values without enforcing this exactly. For total area, the potential is defined as
\begin{align*}
	\Energy_{\text{area}}(f) = \textstyle \big((\sum_{T \in F} a_f(T))/A_0 - 1 \big)^2.
\end{align*}
The raw deviation is normalized by the initial area $A_0$ to make the penalty scale invariant. The total volume potential is defined analogously.

\subsubsection{Static Obstacles}
\label{sec:StaticObstacles}

For practical modeling purposes, it may be desirable not to design an object in isolation, but instead to design it within its intended environment. To that end, we provide the ability to place ``obstacles'', which are static meshes that exert a repulsive force on the optimization surface. These obstacles can be used to model surrounding environments such as rooms and the objects within them, which must be avoided by the object under design. From an obstacle $O$ with embedding $f_O$, each point $x$ in the domain experiences a repulsive potential equal to 
\begin{align*}
	\Energy_{\text{obs}}(x) 
	= 
	\textstyle
	\sum_{S \in F_O} \nabs{f_O(S) - x}^{-p} \, a_{f_O}(S)
\end{align*}
with $p$ matching the exponent of the tangent-point energy. Na\"{i}vely, this requires iteration over all faces of $O$, but Barnes-Hut can be used as in \secref{BarnesHut} to approximate the obstacle potential.

\subsubsection{Implicit Obstacles and Attractors}
\label{sec:ImplicitObstaclesAndAttractors}

Similarly to static mesh obstacles, one can also use implicit surfaces defined by signed distance fields as obstacles or attractors. Given a signed distance field $d : \RR^3 \to \RR$, the repulsive potential experienced at any point $x$ due to the implicit obstacle defined by $d(x) = 0$ is simply
\begin{align*}
	\Energy_{i}(x) = d(x)^{-p}.
\end{align*}
An implicit attractor, rather than repelling other objects away from it, pulls objects towards it. The attractive potential experienced at any point $x$ is simply the reciprocal of the above, or
\begin{align*}
	\Energy_{a}(x) = d(x)^p.
\end{align*}

\subsubsection{Boundary Length and Curvature}
\label{sec:BoundaryLengthAndCurvature}

For meshes with boundary (e.g. \figref{TorusEversion}), it may be beneficial to regularize the shape of the boundary curves. We support two potentials for this purpose. One is a regularizer on the length of the boundary, defined as
\begin{align*}
	\Energy_{b} = \left(L - \textstyle \sum_{e \in \partial M} l(e) \right)^2,
\end{align*}
where $L$ is a target boundary length, and $l(e)$ is the length of boundary edge $e$. The other regularizes the curvature, and is defined as
\begin{align*}
	\Energy_{c} = \textstyle \sum_{v \in \partial M} \theta(v)^2/\ell(v),
\end{align*}
where $\theta(v)$ is the turning angle at vertex \(v\), and \(\ell(v)\) is the dual length (\ie, half the length of the two incident edges).

\subsubsection{Willmore Energy}
\label{sec:WillmoreEnergy}

One can also add surface fairing energies such as the Willmore energy.
For example, we use the following discrete variant of the squared mean curvature integral:
\begin{align*}
	\Energy_{\text{Willmore}}(f)
	 = 
	 f^\T \mathbf{A} \, \mathbf{M}^{-1} \mathbf{A} \, f.
\end{align*}
Here $\mathbf{A} $ is the stiffness matrix of the cotan Laplacian and $\mathbf{M}$ is the lumped mass matrix. Up to mass lumping, this is the discrete Willmore energy from \cite{Dziuk:2008:NUMMA}.
As suggested in \cite{Eckstein:2007:GSF,Schumacher:2017:HGF}, 
we add an $H^2$ inner product term $\mathbf{A} \, \mathbf{M}^{-1} \, \mathbf{A}$ to the matrix that we invert in \secref{PreconditionedIterativeMethod}.

\section{Evaluation and Comparisons}
\label{sec:EvaluationandComparisons}

\begin{figure}
   \includegraphics{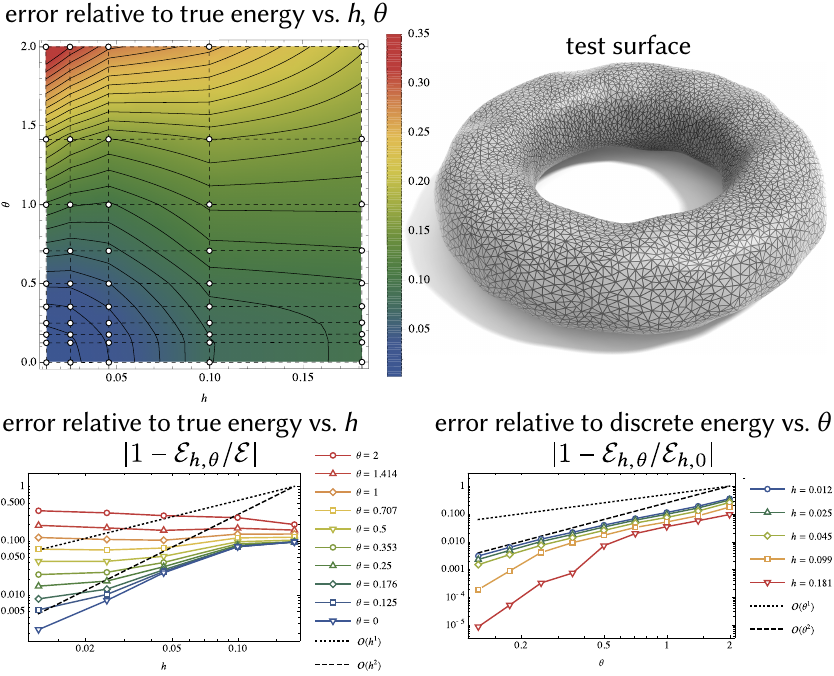}
   \caption{Empirically, our discrete tangent-point energy appears to converge to the true smooth energy at a rate somewhere between \(O(h)\) and \(O(h^2)\); as expected, our Barnes-Hut approximation also converges to the discrete energy as \(\theta \to 0\).  Reference values are obtained by applying highly accurate numerical integration to the tangent-point energy on a smooth parameterized surface (triangulated in \figloc{top right}).  See supplemental for additional examples.\label{fig:ConvergenceStudy}}
\end{figure}

\begin{figure*}[t]
   \includegraphics[width=\textwidth]{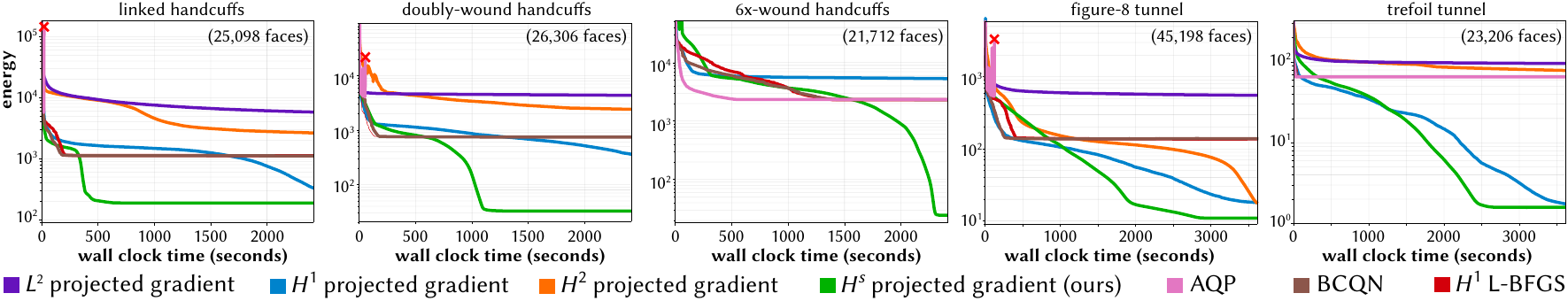}
   \caption{Energy plots showing the effectiveness of a suite of methods at minimizing the tangent-point energy. Our $H^s$ method (in green) reaches minimizers more quickly and consistently than the alternatives. Points at which methods became unstable are marked with an X. Renderings of the meshes used and their minimizers can be seen in \figref{IsotopyGallery}.\label{fig:PerformanceGraphs}}
\end{figure*}

\subsection{Consistency Testing}

Evaluating convergence of our discretization and approximation scheme to minimizers is not straightforward, since to date there are only conjectures about what minimal solutions might look like (\secref{CanonicalEmbeddings}).  Instead, we numerically investigate the \emph{consistency} of our energy discretization:  We generate several smooth surfaces, compute their true tangent-point energies, and compare to our discrete energy and its Barnes-Hut approximation. 

The exact energy can be computed directly only for very simple shapes, like a round sphere or torus of revolution.
To get a more generic picture, we took the parameterized torus of
revolution $f_0(\phi,\theta) = \big( (1 + \frac{1}{3} \cos(\phi)) \, \cos(\phi), (1 + \tfrac{1}{3} \cos(\theta)) \, \sin(\phi), \frac{1}{3} \, \sin(\theta)\big)$ and perturbed it by a random trigonometric polynomial $\Phi \colon \RR^3 \to \RR^3$ of small magnitude (to ensure embeddedness) and small order (to obtain moderate curvature) to obtain the final smooth surface $f := f_0 + \Phi \circ f_0$.
We computed $\EnergyP(f)$ up to $6$ digits of precision by numerical integration with \emph{Mathematica}'s \texttt{NIntegrate} command using the \texttt{"LocalAdaptive"} strategy.
Afterwards, we computed an affinely squeezed Delaunay triangulation of  $[0,2 \, \pi] \times [0, 2 \,\uppi]$ and used it to sample the surface $f$. 
The remaining nonuniformities in triangle size and aspect ratio were repaired by the remeshing routine from \secref{DynamicRemeshing} followed by projecting each resulting vertex position back to the surface $f$. For the resulting discrete surface $f_h$ we computed its Barnes-Hut energy $\ApproxDiscEnergyP(f_h)$ (see \eqref{BarnesHutEnergy}) for various values of the separation parameter $\theta$; in the case $\theta = 0$, this is the all-pairs energy $\DiscEnergy(f_h)$ from \eqref{DiscreteEnergy}.
The resulting relative errors are shown in \figref{ConvergenceStudy}.

The discrete energy $\DiscEnergyP(f_h)$ employs the face normals, which are known to be consistent of order $1$ only. That means, their error is $O(h)$, where $h>0$ denotes the longest edge length.
So it is expected that the discretization error $e_{h} := |\DiscEnergyP(f_h) - \EnergyP(f)|$ is no better than $O(h)$. Surprisingly, the experiments show that the numerical rate is considerably better (see \figref{ConvergenceStudy}, \figloc{bottom left} and for $\theta = 0$).
Moreover, we use center of mass data on BVH nodes; 
so the deviation $e_{h,\theta} :=  |\ApproxDiscEnergyP(f_h) - \DiscEnergyP(f_h)|$ of the Barnes-Hut approximation from the discrete energy should be dominated by the midpoint rule's consistency error which is $O(\theta^2)$. Indeed our experiments seem confirm this (see \figref{ConvergenceStudy}, \figloc{bottom right}). 

\subsection{Comparison of Optimization Methods}

We next compare to other accelerated descent strategies from geometry processing and geometric optimization.  Our overall observations are consistent with those from \citet{Yu:2021:TOG}: the fractional Sobolev scheme converges to local minimizers far quicker than general-purpose acceleration strategies (dramatically so, in the case of highly knotted configurations).  This should not come as a surprise: the all-pairs energy we seek to minimize behaves very differently from those arising in, \eg{}, curvature flows or elasticity, which are based on discrete differential operators with small local stencils.

To make a fair comparison, all methods use identical code for accelerated energy and differential evaluations (\secref{BarnesHut}), and differ only in how they use these values. The same dynamic remeshing routine (\secref{DynamicRemeshing}) is also run at the end of each iteration for all methods.  Note that edge splits and collapses invalidate the history of methods such as L-BFGS; here we use memory vectors for as long as they are valid, and reset them when edge splits or collapses occur.  All experiments were run with barycenter and total area constraints. Since AQP and L-BFGS methods do not support nonlinear constraints such as total area---for these methods, we instead use stiff penalty functions (\secref{TotalAreaAndVolumePotentials}) to discourage excessive drift.

\paragraph{Comparison Methods.} Our comparisons are guided by the extensive comparisons carried out in \citet[Section 7]{Yu:2021:TOG}; here we compare with the best of those methods.  As a baseline we consider ordinary \(L^2\) gradient descent, which amounts to replacing \(A\) in \eqref{SaddleProblem} with the mass matrix.  Likewise, replacing \(A\) with the weak Laplacian \(\Delta\) (encoded by the cotan matrix) yields standard \(H^1\) Sobolev preconditioning; \(H^2\) Sobolev preconditioning is achieved by solving \eqref{SaddleProblem} with the weak formulation of the bi-Laplacian \(\Delta^2\) in place of \(A\).  (This latter preconditioner is essentially an ideal choice for Willmore flow~\cite{Schumacher:2017:HGF}.) Like $H^1$ preconditioning, the \emph{accelerated quadratic proxy (AQP)} method uses the weak Laplacian $\Delta$ as the inner product, but also computes a Nesterov acceleration step from the previous two configurations; this strategy is compatible only with linear constraints~\cite[Section 2]{Kovalsky:2016}.  Another common strategy, which we refer to as \(H^1\) L-BFGS, is to initialize L-BFGS with the weak Laplacian rather than the identity matrix, and likewise use the Laplacian to evaluate inner products.  Finally, \emph{Blended cured quasi-Newton (BCQN)} essentially interpolates between ordinary \(H^1\) Sobolev preconditioning and \(H^1\) L-BFGS, together with barrier penalties to prevent triangle inversion.  Since our gradient is almost \(H^s\) orthogonal with tangential motions of the surface (and do not experience element inversions), we omit these penalties.

\subsection{Time Step Restriction}
\label{sec:TimeStepRestriction}

\figref{BumpyMug} verifies that matching the order of the inner product to that of the energy differential essentially lifts the mesh-dependent time step restriction. Here, we sampled the same surface at three resolutions, and ran each method for the same number of iterations.  Our \(H^s\) scheme makes more progress for an equal number of iterations---but more importantly, the per-iteration progress of \(H^s\) is largely unaffected by mesh resolution, whereas all other methods slow down as resolution increases.  Hence, even if some of these methods could be further accelerated by a constant factor (\eg{}, via code-level optimization), asymptotic behavior would ultimately dominate.

\begin{figure*}[t]
   \includegraphics[width=\textwidth]{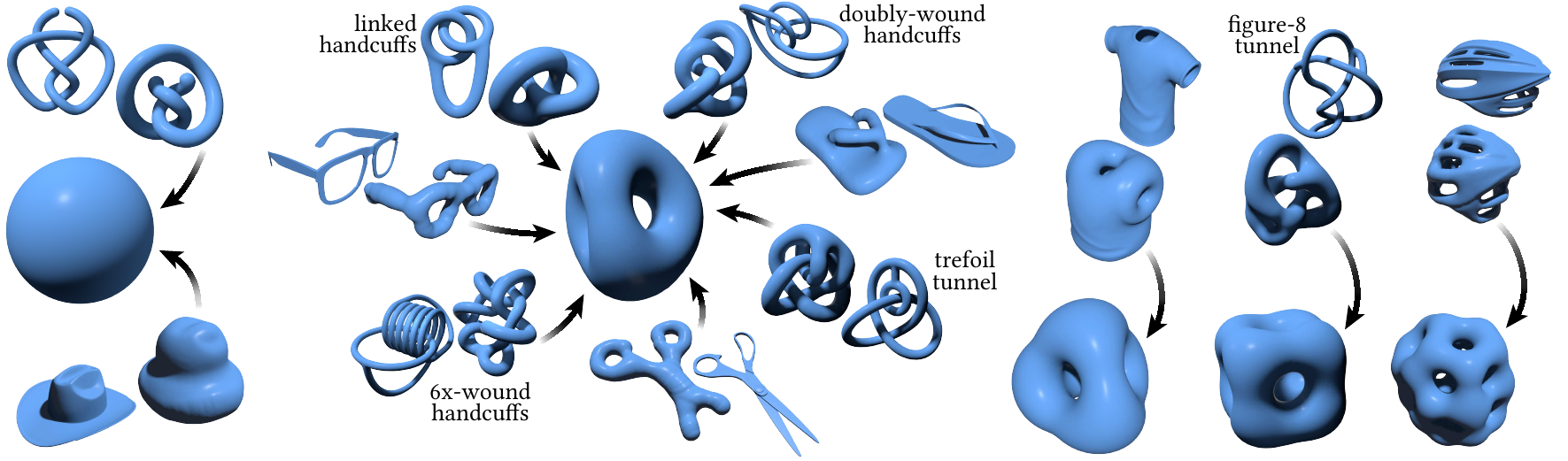}
   \vspace{-1.5\baselineskip}\caption{Gallery of isotopies obtained by minimizing tangent-point energy---notice that highly knotted surfaces, as well as surfaces with thin sheets and handles, successfully flow to their canonical embeddings. Surfaces are grouped by their isotopy equivalence classes, which are extremely difficult to determine via visual inspection (and also not simply determined by Euler characteristic---see \figref{Genus2Table}).  Labeled meshes are used for performance comparisons in \figref{PerformanceGraphs}. \label{fig:IsotopyGallery}}
\end{figure*}

\subsection{Wall-Clock Performance}
\label{sec:Wall-ClockPerformance}

We also timed the real-world performance of each method on several challenge meshes, using an AMD Ryzen Threadripper 3990X with 32 GB of RAM.
Though in practice our solver benefits from multiple threads (see \secref{Introduction}), we ran this benchmark single-threaded to ensure a fair comparison.  \figref{PerformanceGraphs} plots energy as a function of time; we ran each method for 3600 seconds for the figure-8 and trefoil tunnels, and 2400 seconds for all others. Reference energy values were computed by evaluating the exact energy, without Barnes-Hut approximation.  Our $H^s$ projected gradient method gave the best performance in all cases, reliably reaching a minimum within the alloted time. In some cases the initial rate of decrease is faster for other methods, likely because there are initially many small local features to be smoothed out.  Subsequently, however, these methods make much slower progress at evolving the global shape.  Though AQP and BQN are also based on \(H^1\) preconditioning, they do not do as well here as the ``vanilla'' \(H^1\) preconditioner.  One possible reason is that these methods do not support hard nonlinear constraints, and hence penalty forces may fight with the main objective.  See \citet{Yu:2021:TOG} for much more extensive discussion and analysis of fractional methods versus a similar set of alternatives.

\section{Examples and Applications}
\label{sec:ExamplesandApplications}

We here explore a variety of applications that help to further evaluate our method, show how it can be used in context, and also identify issues that might be improved in future work.  These applications are also illustrated in the accompanying video---note that for many of these examples we take time steps far smaller than the optimal step determined by line search, in order to produce smooth animation.

\subsection{Mathematical Visualization and Exploration}
\label{sec:MathematicalVisualization}

Mathematically, the motions computed by our method are ambient isotopies: given two embeddings \(f_0, f_1: M \to \mathbb{R}^3\), an \emph{ambient isotopy} is a continuous map \(F: \mathbb{R}^3 \times [0,1] \to \mathbb{R}^3\) such that for all \(x \in M\), \(F(x,0) = x\), \(F(f_0(x),1) = f_1(x)\), and \(F(x,t)\) is a homeomorphism from \(\mathbb{R}^3\) to \(\mathbb{R}^3\) for every time \(0 \leq t \leq 1\).  
Intuitively, an ambient isotopy is a deformation of space that ``drags along'' \(f_0\) with it, turning it into \(f_1\) while avoiding any changes to the initial topology.  A basic question in geometric topology is whether two embedded manifolds are ambiently isotopic, and in general this question can be quite hard to answer---for instance, even detecting whether an embedding of the circle in \(\mathbb{R}^3\) is equivalent to the unit circle (or ``unknot'') has not yet admitted a polynomial time algorithm~\cite{Lackenby:2016:EKT}.  Hence, computational tools have been developed to explore such questions experimentally, with a notable example being the widely-used \emph{KnotPlot} package for curve untangling~\cite{Scharein:1998:ITD}.  The software developed for our project effectively provides the first ``KnotPlot for surfaces.''  Especially the fact that our solver exhibits rapid convergence and excellent scaling enables us to investigate questions that would be impossible with na\"{i}ve numerical methods.

\begin{figure}
   \includegraphics[width=\columnwidth]{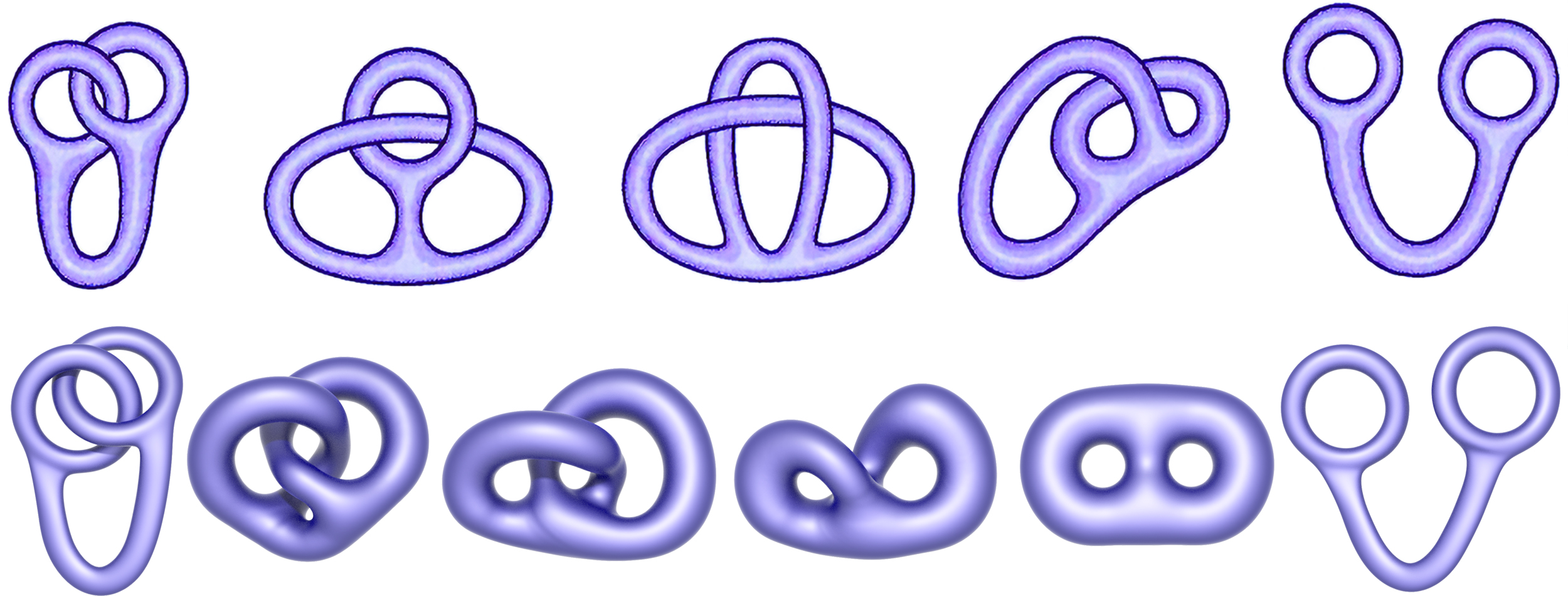}
   \caption{\figloc{Top:} even careful illustrations of topological phenomena (here drawn by mathematician Peter Lynch) can be difficult to understand without a good visual imagination. \figloc{Bottom:} our method automatically generates continuous motions that are easier to interpret (see video), enabling exploration by students and researchers who do not have significant artistic training.\label{fig:HandcuffsDetail}}
\end{figure}

\begin{figure}
   \includegraphics[width=\columnwidth]{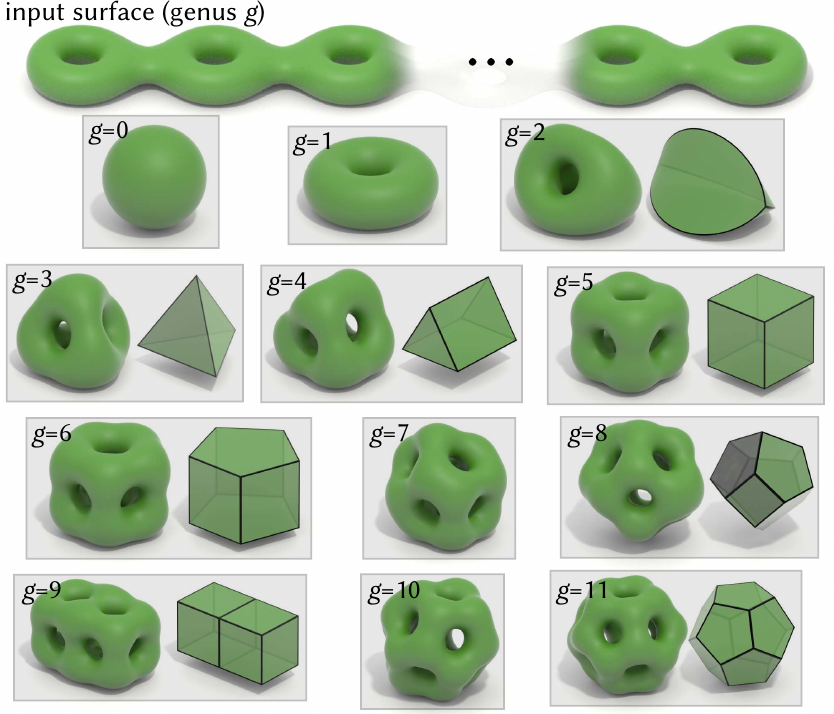}
   \caption{Global minimizers of geometric energies provide canonical domains that can be used to map between surfaces of the same topology, or simply help visualize a topological space.  Here we show conjectured minimizers of tangent-point energy for unknotted surfaces of genus \(g\); adjacent figures illustrate symmetries (when present).\label{fig:GenusGMinimizers}}
\end{figure}

\subsubsection{Canonical Embeddings}
\label{sec:CanonicalEmbeddings}

Global minimizers of geometric energies provide the ``simplest'' possible geometric representative of a given topological space.  Such minimizers also play a critical role in geometric algorithms since they provide a canonical domain for, \eg{}, surface correspondence and data transfer---see for instance recent algorithms in both the intrinsic~\cite{Schmidt:2020:ISM,Gillespie:2021:DCE} and extrinsic~\cite{Kazhdan:2012:CMC,Ye:2018:UDF} settings.  Formally proving that a given surface is a global minimizer is quite challenging.  For instance, even the classic \emph{Willmore conjecture} (which says that the \emph{Clifford torus} minimizes Willmore energy for genus-1 surfaces) was resolved only very recently, after about 50 years of sustained effort~\cite{Marques:2014:MMT}.  Hence, numerical tools are essential for formulating hypotheses about the behavior of minimizers and other critical points.  To date, there are no clear conjectures about tangent-point minimizers for surfaces of genus \(g \geq 2\).  For reasons discussed in \secref{Introduction}, these minimizers likely exhibit symmetries in \(\mathbb{R}^3\) rather than \(\mathbb{S}^3\), making them potentially useful as a base domain for algorithms in extrinsic shape processing.  To do so, one would simply need to track the parametric correspondence (\eg{}, via UV-coordinates), and perhaps minimize tangential distortion after flowing to a geometric minimizer (\ala{} \citet{Schmidt:2020:ISM}).

\paragraph{Unknotted Minimizers} \figref{GenusGMinimizers} shows a numerical study for untangled surfaces of increasing genus, initialized with a linear arrangement of handles.  For genus 0, 1, and 2 we get a round sphere, a torus of revolution, and a surface with symmetries of a triangular prism.  Other surfaces appear to exhibit symmetries of a highly regular polyhedron---for instance, for genus 3, 4, 5, 6, 8, 9, and 11 we get symmetries of the tetrahedron, triangular prism, cube, pentagonal prism, truncated bipyramid, rectangular prism, and dodecahedron, respectively.  Symmetries (if any) for genus 7 and 10 are less clear---or we may have simply failed to reach a global minimum.  Interestingly, an octahedral configuration does not appear to be a minimizer for genus 7, even if we start with a symmetric configuration (and similarly for the icosahedron, not shown).  In general it seems that triangular ``faces'' are not preferred in higher-genus configurations due to the small angle between ``edges''---much as electron repulsion maximizes bond angles in molecular geometries (\eg{}, stable compounds like graphite prefer bond angles near \(120^\circ\), whereas only unstable compounds like white phosphorus exhibit tetrahedral symmetry).

\begin{figure}
   \includegraphics[width=\columnwidth]{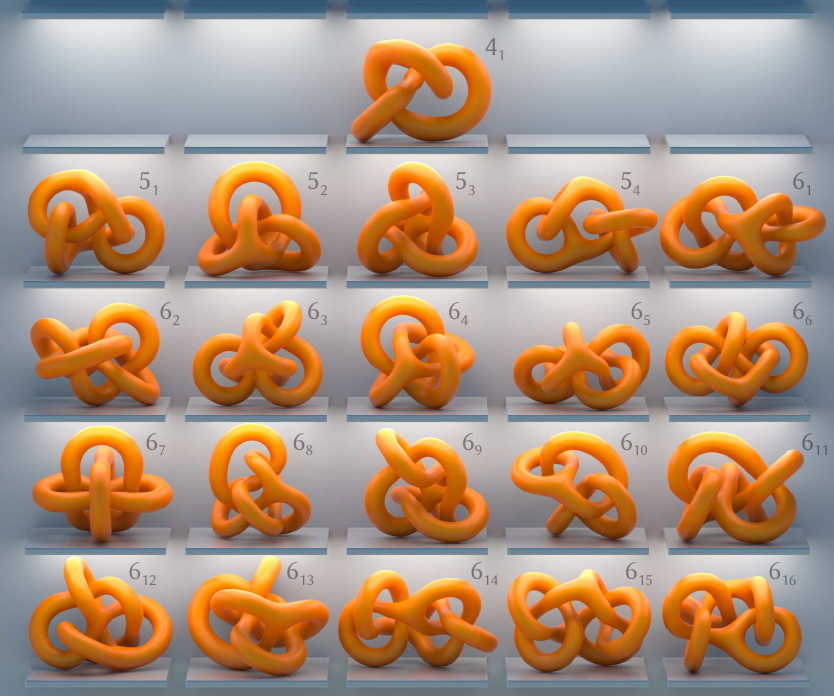}
   \caption{Geometric functionals provide a bridge between topology and geometry by enabling one to construct canonical geometric representatives of a given topological space.  Here, minimizers of tangent point energy are used to visualize nontrivial isotopy classes of a genus-2 surface.  (Numbers indicate number of crossings; subscripts index trivalent graphs from \cite[Table 1]{Ishii:2012:TGT}).\label{fig:Genus2Table}}
\end{figure}

\setlength{\columnsep}{1em}
\setlength{\intextsep}{0em}
\begin{wrapfigure}{r}{61pt}
   \includegraphics{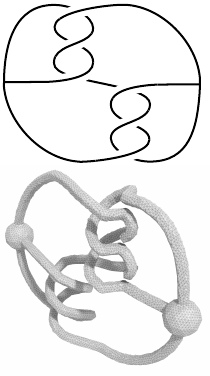}
\end{wrapfigure}
\paragraph{Knotted Minimizers} A key feature of tangent-point energy (versus, say, Willmore energy) is that it enables us to find minimizers within a given isotopy class.  Hence, just as it is quite common to make tables of canonical knot embeddings, we can now make tables of canonical embeddings for knotted surfaces.  For instance, \figref{Genus2Table} shows the first-ever visualization of the different ways a genus-2 surface can be embedded in space.  In the past, these isotopy classes have been depicted only as trivalent graphs---we take each such graph from \cite[Table 1]{Ishii:2012:TGT}, and construct a topologically equivalent initial mesh that is optimized by our approach (see inset).  As with knots most of these minimizers do not exhibit much extrinsic symmetry, except for, \eg{}, \(6_7\) and \(5_3\) which exhibit bilateral and 3-fold symmetry, \resp{}

\begin{figure}
   \includegraphics[width=\columnwidth]{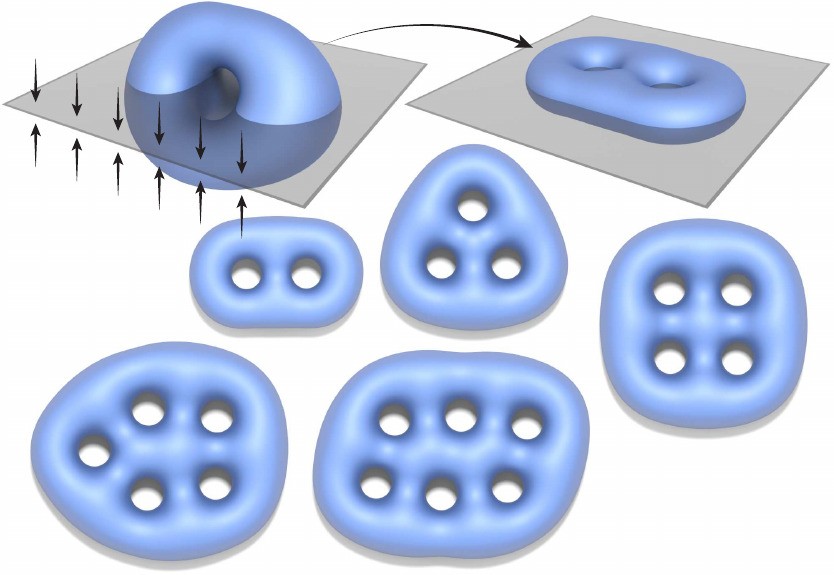}
   \caption{\figloc{Top:} minimizers of tangent-point energy often exhibit three-dimensional symmetries which can be difficult to understand from a single view---by adding an attracting plane, we get embeddings that can be nicely displayed in two-dimensional illustrations. \figloc{Bottom:} constrained minimizers for genus 2 through 6.\label{fig:FlatteningEmbedding}}
\end{figure}

\setlength{\columnsep}{1em}
\setlength{\intextsep}{0em}
\begin{wrapfigure}{r}{100pt}
   \includegraphics[width=100pt]{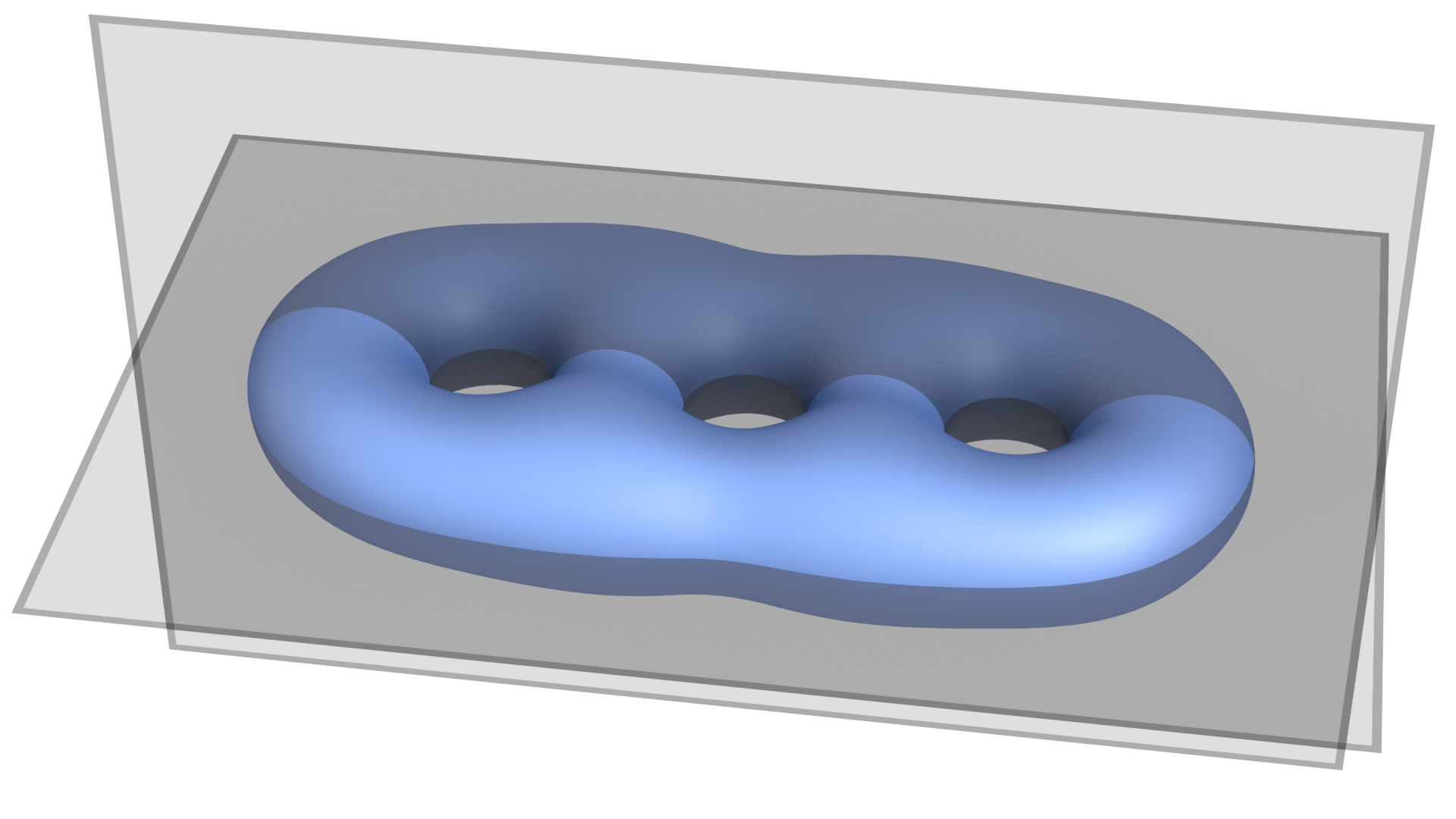}
\end{wrapfigure}
\paragraph{Planar Representatives.} Although minimizers exhibit a high degree of symmetry in \(\mathbb{R}^3\), it can be hard to determine even the genus of a minimizer when viewed from just a single viewpoint.  In contrast, topological figures depicted by expert illustrators tend to be somewhat ``2.5-dimensional'' so that they can be better understood when projected onto the image plane.  We can replicate this behavior by adding a simple attractive plane potential, as depicted in \figref{FlatteningEmbedding}, yielding minimizers that are much easier to recognize (contrast with \figref{GenusGMinimizers}).  An additional plane constraint yields a linear arrangement of handles, as commonly drawn by hand (see inset).

\subsubsection{Illustrating Isotopies}
\label{sec:IllustratingIsotopies}

Our method also provides significant utility for mathematical visualization and illustration.  Traditionally, interesting homotopies and isotopies are depicted by a sequence of drawings (or perhaps physical models) highlighting key moments of transition---a practice that has developed over time into a true art form~\cite{Francis:1987:TP}.  However, even the best drawings can be difficult to understand without significant thought and visual imagination.  To obtain continuous motions (that are more easily understood), a small number of carefully ``hand-crafted'' computer animations have been produced over the years by either artist keyframing, or explicit programming of meticulously derived parametric formulas~\cite{Levy:1995:MW,Bednorz:2019:ASE}.  More recently, automatic optimization-based tools have been used to produce animations, such as the \emph{minimax sphere eversion}~\cite{Francis:1997:MSE}, as well as recent work in computer graphics on metric embedding~\cite{Chern:2018:SFM} and conformally-constrained Willmore surfaces~\cite{Soliman:2021:CWS}.  Since these optimization-based tools are largely automatic, they help to democratize the creation of topological animations---our scheme extends such tools to the important and difficult case of ambient isotopies.

\begin{figure}
   \includegraphics[width=\columnwidth]{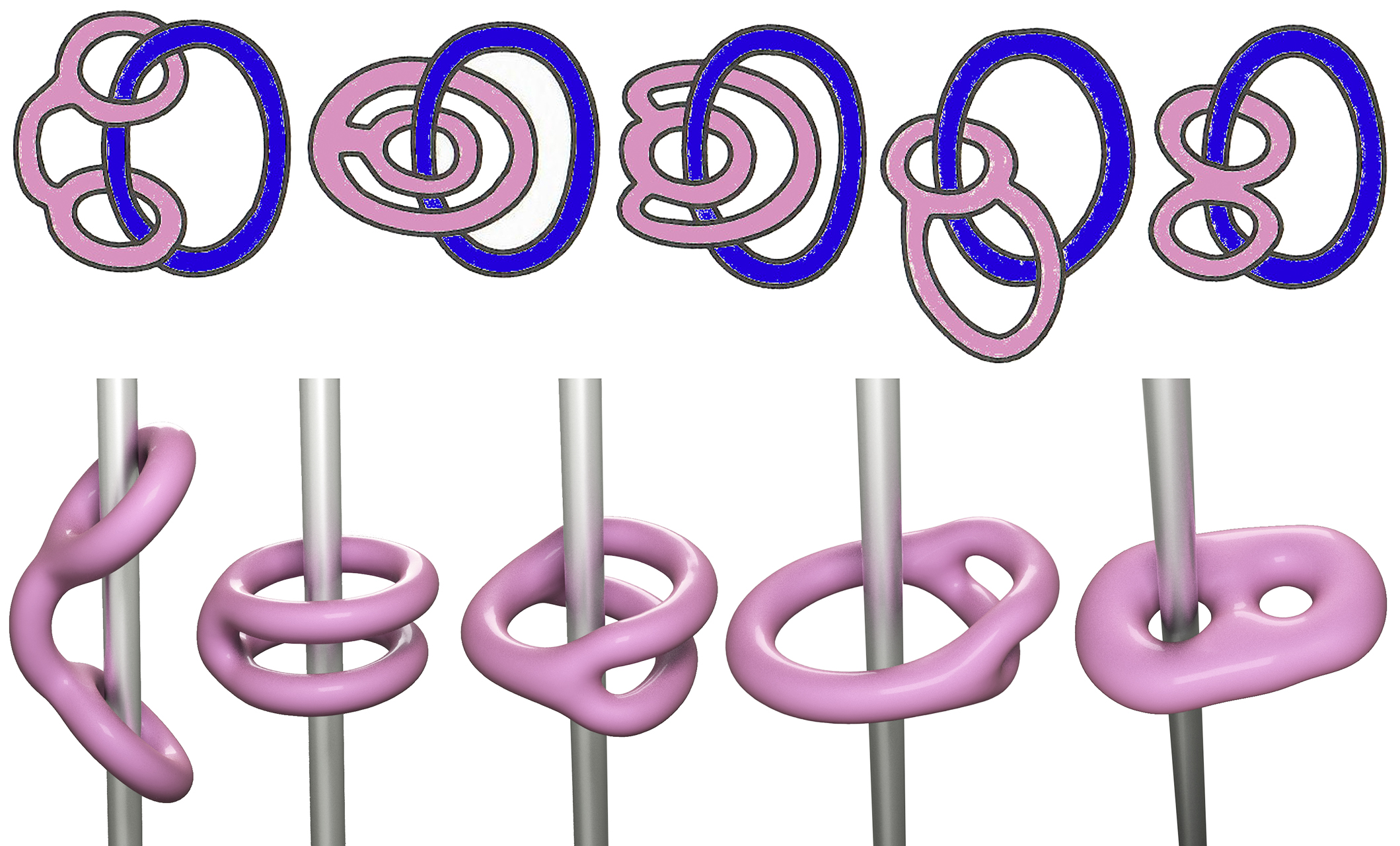}
   \caption{Surprisingly, one can remove a handle of a double torus from a loop or pole without cutting or pinching the surface.  \figloc{Top:} hand-drawn illustration by \citet{Wells:1997:PDC}. \figloc{Bottom:} isotopy computed automatically by our method (see video); no keyframing or boundary conditions were used.\label{fig:HandcuffsPole}}
\end{figure}

\setlength{\columnsep}{1em}
\setlength{\intextsep}{0em}
\begin{wrapfigure}{r}{118pt}
   \includegraphics{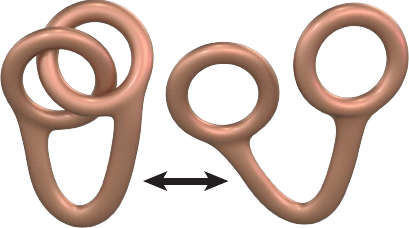}
\end{wrapfigure}
One classic example is ``unlinking'' a pair of handcuffs (as shown in the inset), though mathematically speaking these handcuffs are not actually linked: surprisingly, they belong to the same isotopy class.  
\figref{HandcuffsDetail} compares a hand drawing of this isotopy with a different isotopy automatically computed via our method---and which is much better depicted in the accompanying video.  
To create this animation we simply minimize tangent-point energy from both start and end configurations, together with a potential that encourages the surface to lay parallel to the view plane.  
Since we reach the same minimizer in both cases (seen in \figref{Teaser}, \figloc{far right}), we can compose these two sequences (one in reverse) to depict the complete motion.  
Other similar examples are shown in \figref{IsotopyGallery}, and in the video.

\figref{HandcuffsPole} shows another classic example: removing one handle of a pair of handcuffs from a rigid pole or ring.  The hand-drawn illustration helps to indicate several stages of this isotopy, which are also captured in our animation.  However, the remarkable fact about our version is that it is driven \emph{purely} by energy minimization---we did not perform any keyframing, nor impose any boundary conditions, yet it still constructs an isotopy in several ``stages'': flatten
\setlength{\columnsep}{1em}
\setlength{\intextsep}{0em}
\begin{wrapfigure}{r}{69pt}
   \includegraphics{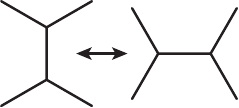}
   \vspace{-.5\baselineskip}\caption{An IH-move.\label{fig:IHMove}}
\end{wrapfigure}
\noindent out the two handles, perform a so-called \emph{IH-move} (see \figref{IHMove} and \cite{Ishii:2008:MIK}), and then optimize the geometry of the untangled surface.  Our specific setup here is to minimize tangent-point energy while fixing surface area, and incorporating an infinite repulsive cylinder (modeled by an implicit surface). As in the previous example we use an attractive plane orthogonal to the pole to obtain a more canonical-looking minimizer.  The only hand-tuning was reducing the repulsive strength of the cylinder near the end of the animation, to give the handles of the final surface a similar size.  Importantly, allowing the barycenter to float freely (\ala{} \secref{FixedBarycenterConstraint}) is essential here, since the center of mass must ultimately move away from the pole.

\begin{figure}
   \includegraphics{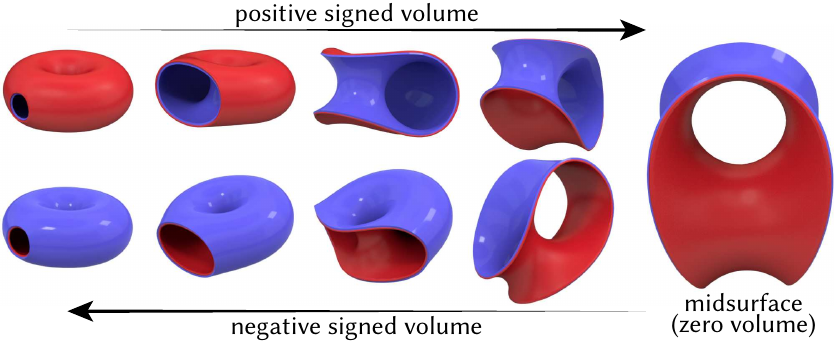}
   \caption{Minimizing the tangent-point energy of a punctured torus while pushing signed volume toward zero yields a surface with reflection symmetry. Applying a reflection and reversing the flow hence yields an eversion that turns the surface ``inside-out'' while avoiding self-intersections.\label{fig:TorusEversion}}
\end{figure}

\paragraph{Punctured Torus Eversion.} Our discrete tangent-point energy can also be evaluated on surfaces with boundary, since we simply take a sum over pairs of triangles.  Since we did not develop a careful treatment of boundary conditions, we simply penalize the total length and total squared curvature to ensure the boundary at least remains regular.  In \figref{TorusEversion} we use this setup to compute an isotopic \emph{eversion} between the two orientations of a punctured torus.  Unlike the classical sphere eversion, where one typically starts with a symmetric midsurface and flows toward the round sphere, we start with the punctured torus and use our flow to find the mid-surface.  The key observation is that the \emph{oriented} volume of the surface \(\int_M \ninnerprod{f(x),N_f(x)}\ dx_f\) will be zero for a symmetric configuration; fixing the area ensures that our zero-volume penalty does not cause the surface to collapse to a point.  Once we reach zero volume we transform the midsurface by a reflection and 90-degree rotation, and run the same flow in reverse (with opposite colors) to obtain the eversion.

\subsection{Geometry Processing and Shape Modeling}
\label{sec:GeometryProcessing}

The no-collision condition is also natural in geometry processing and shape modeling, especially when a surface is meant to represent the boundary of a solid object (\eg{}, for computational fabrication).  As noted in \secref{Introduction}, there has been relatively little work on collision-aware geometric modeling---see for instance \citet{Harmon:2011:IAG} and references therein.  In contrast to resolving local intersections, tangent-point energy adds the complementary functionality of global collision avoidance to a broad range of existing tasks.  Here we present several aspirational examples---importantly, our goal is not to outperform more specialized, mature solutions, but rather to explore how a tangent-point regularizer might serve as a unified approach to collision avoidance across many disparate applications.

\setlength{\columnsep}{1em}
\setlength{\intextsep}{0em}
\begin{wrapfigure}{r}{75pt}
   \includegraphics{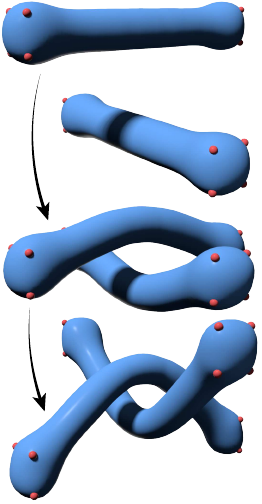}
\end{wrapfigure}
\newpage

\begin{figure}
   \includegraphics[width=\columnwidth]{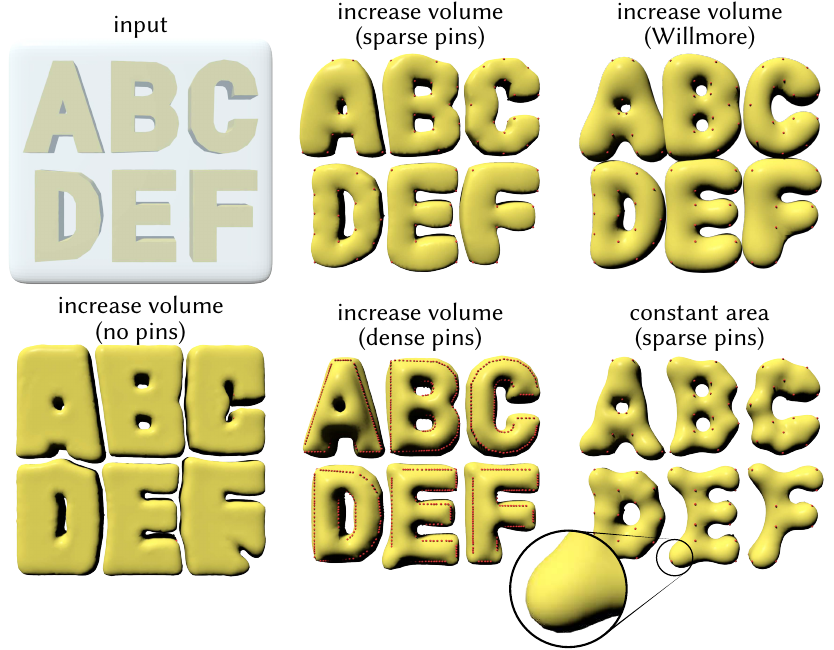}
   \caption{The tangent-point energy can be used to make variational surface modeling responsive to proximity, rather than just collisions.  Here for instance we pin a sparse or dense set of points and modify volume and surface area to adjust the appearance of some text (in some cases enclosed in a box).  Like Willmore energy \figloc{(top right)}, we get smooth behavior near point constraints (see magnified portion), but avoid overlap.\label{fig:ABCDEF}}
\end{figure}

\paragraph{Proximity-Aware Variational Modeling} As a basic example, the inset figure above shows a simple example of interactive surface editing, where surface geometry is guided by point constraints, and nearby geometry is moved out of the way by the tangent-point energy.  To better preserve the details of an initial mesh one might also combine tangent-point energy with a discrete shell energy~\cite{Grinspun:2003:DS}, which would entail transferring the material configuration across meshing operations (a question which is beyond the scope of this work).  \figref{ABCDEF} shows another example where pinned points and edges are interpolated while optimizing the rest of the geometry.  (Here we disable remeshing, but could easily modify remeshing to ignore pinned vertices).  Unlike harmonic interpolation or area minimization, for which point constraints are ill-posed, we get nice curvature behavior even near the pins; unlike Willmore flow (which provides good curvature behavior), we avoid self-intersection.  Tangent-point energy could also in principle be used as a regularizer to discourage collision in other common modeling paradigms, such as \emph{as rigid as possible (ARAP)} modeling~\cite{Sorkine:2007:ARA}.

\begin{figure}
   \includegraphics[width=\columnwidth]{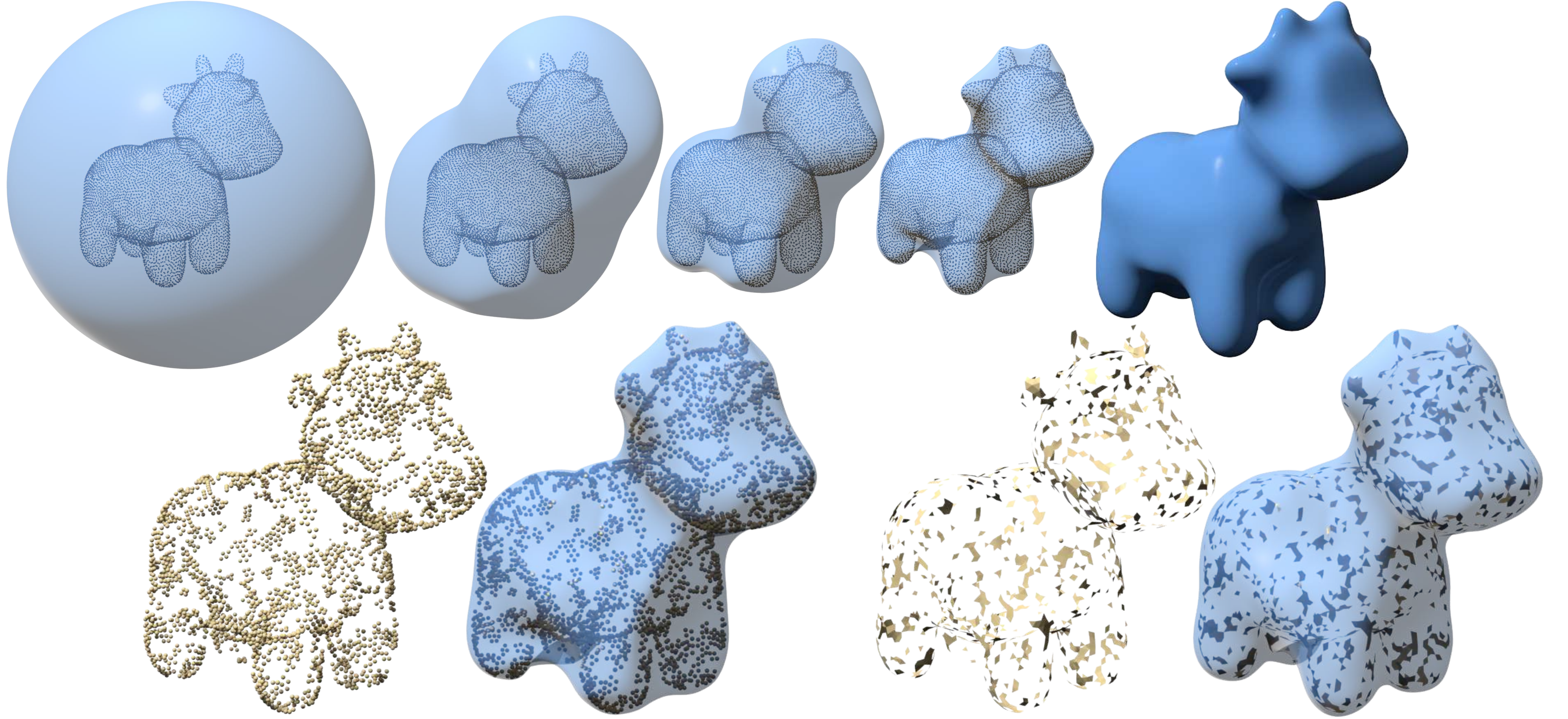}
   \caption{Here we perform a simple ``shrink wrapping'' to obtain a manifold, intersection-free reconstruction \figloc{(top}), which works well even for points or polygon soup with severe holes and missing data.\label{fig:PointCloudWrapping}}
\end{figure}

\subsubsection{Shrink Wrapping}
\label{sec:ShrinkWrapping}

One class of methods for reconstructing a surface from a collection of points is to ``shrink-wrap'' them with a triangle mesh~\cite{Kobbelt:1999:SWA,Hanocka:2020:P2M}; such methods are especially suitable in problems where one wishes to fit a high-quality template mesh to a known class of shapes (\eg{}, head or body scans).  A basic problem, however, is that the mesh can get ``tangled'' during wrapping, inhibiting progress or requiring intricate remeshing to resolve self-intersections.  Tangent-point energy may prove useful as a regularizer for such methods---\figref{PointCloudWrapping} shows a basic shrink wrapping example on a point cloud, and on polygon soup with severe holes. Here we minimize tangent-point energy with a gradually decreasing volume constraint.

\subsubsection{Nested Envelopes}
\label{sec:NestedEnvelopes}

\begin{figure}
   \includegraphics[width=\columnwidth]{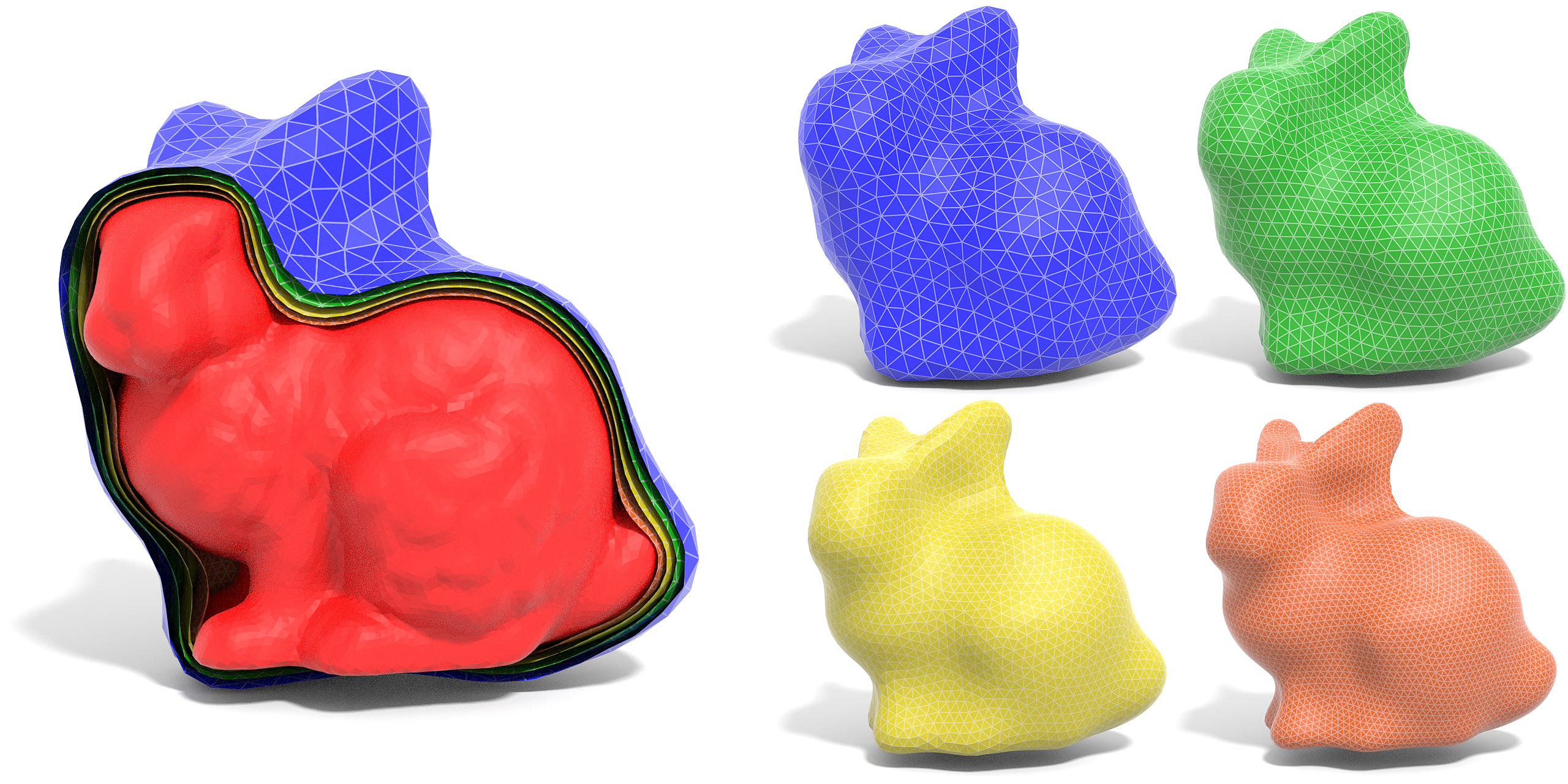}
   \caption{We can ``shrink wrap'' a model to get a sequence of progressively coarser approximating envelopes that exhibit a strict containment property, and are free of self-intersection.  Here we aim for a 1.15x increase in volume at each level.\label{fig:Nested}}
\end{figure}

In a similar vein, nested sequences of solids \(U_1 \subset \cdots \subset U_k \subset \mathbb{R}^\DomDim\) represented by progressively coarser meshes have applications in multiresolution solvers, cage-based editing, and physical simulation~\cite{Sacht:2015:NC}.  In \figref{Nested} we construct each surface \(\partial U_k\) by minimizing tangent point energy plus a volume constraint, and gradually adjusting the constrained volume to achieve a fixed constant factor (here, 1.15x) of the volume of \(\partial U_{k-1}\).  This variational approach may offer interesting generalizations of ordinary nested cages, since it can easily incorporate constraints and objectives beyond just collision avoidance.

\begin{figure}
   \includegraphics[width=\columnwidth]{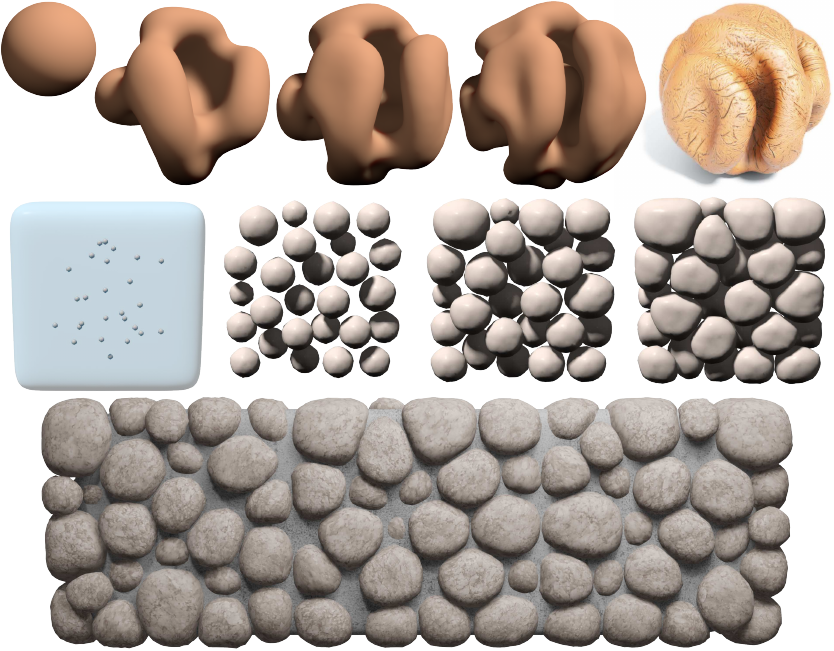}
   \caption{We can also use tangent-point energy for generative modeling by ``growing'' a surface subject to constraints.  \figloc{Top:} confining to a sphere while increasing area leads to a wrinkled shape reminiscent of a walnut. \figloc{Bottom:} growing many small spheres inside a slab yields a tileable cobblestone pattern.\label{fig:GenerativeModeling}}
\end{figure}

\subsubsection{Generative Modeling}
\label{sec:GenerativeModeling}

Rather than using the tangent-point energy to edit or process existing data, we can also use it to generate new geometry.  In nature, the growth of organic shapes is often governed by simple combinations of objectives, \eg{}, a balance between area and volume while avoiding self-collision.  We can likewise use such forces to drive the growth of organic-looking objects, such as the ``walnut'' depicted in \figref{GenerativeModeling}, \figloc{top}.  The same technique is used in \figref{GenerativeModeling}, \figloc{bottom}, where multiple objects are packed into a volume to create a repeating organic pattern.

\subsubsection{Collision Resolution}
\label{sec:CollisionResolution}

\begin{figure}
   \includegraphics[width=\columnwidth]{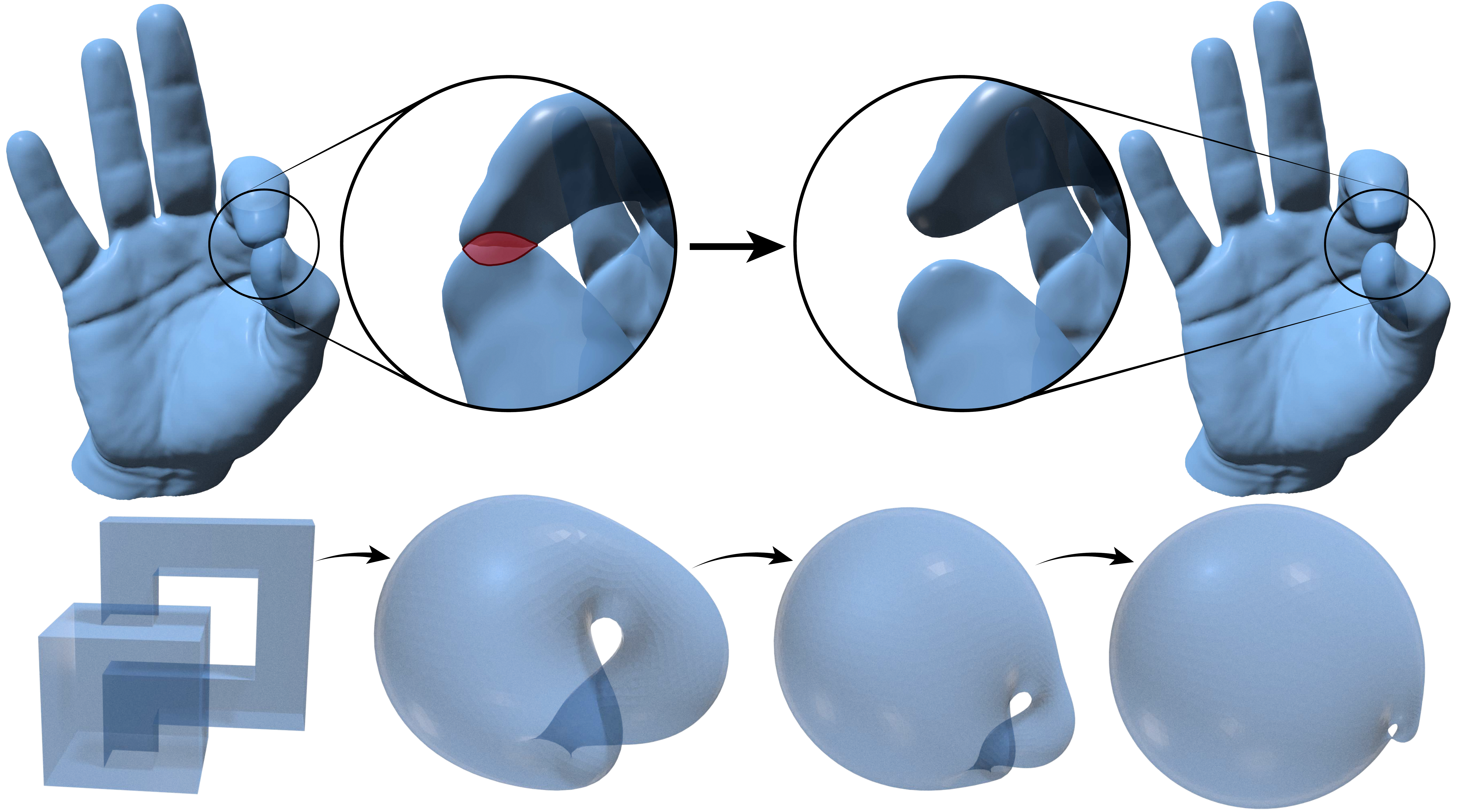}
   \caption{For exponents \(p<4\), the tangent-point energy \(\EnergyP\) is no longer infinite for self-intersecting surfaces, but still discourages overlap.  Here we try using this ``subcritical'' energy to resolve intersections, which works for small intersections \figloc{(top)}, but fails for an unembeddable surface like the Klein bottle \figloc{(bottom)}.\label{fig:CollisionResolution}}
\end{figure}

In many geometry processing tasks, input data is not free of self-intersections.  For exponents $p > 4$, the tangent-point energy $\EnergyP$ of a non-embedded surface is infinite; to \emph{resolve} intersections in the input, we can try reducing the exponent to a value \(p < 4\), at which point \(\EnergyP\) becomes finite but still discourages collision.  Here we find that it also helps to disable the low-order term from \eqref{LowOrderTerm}. Empirically, the same system framework now appears capable of eliminating small self-intersections (\figref{CollisionResolution}, \figloc{top}), through struggles in more difficult scenarios like the Klein bottle depicted in \figref{CollisionResolution}, \figloc{bottom}, which cannot be globally embedded without self-intersection.  Further analysis of the energy for these ``subcritical'' values may help to provide more robust tools for global collision resolution.

\section{Limitations and Future Work}
\label{sec:LimitationsandFutureWork}

The experiments from \secref{ExamplesandApplications} suggest many opportunities for improvement.  For instance, significant performance gains could be achieved purely through better software engineering, \eg{}, improving our parallel implementation of hierarchical matrix multiplication (which is currently bottlenecked around 4--8 threads), or implementing curvature-adaptive remeshing (\ala{} \cite{Dunyach:2013:ARR}), rather than finely tessellating the whole domain.  It would also be quite useful to track mesh attributes across remeshing operations, to enable (for instance) mapping of data from one shape to another through the canonical minimizer.  Since we discretize tangent-point energy, we can provide no formal guarantee that collisions will not occur---as in \citet[5.4]{Yu:2021:TOG}, a pragmatic solution would be to use continuous-time collision detection to limit the time step (or simply provide a certificate).

Several issues require deeper investigation.  For one thing, unlike \citet{Yu:2021:TOG}, our preconditioning strategy cannot easily accommodate dense constraints (\eg{}, preservation of each triangle area), which would require a prohibitive number of iterative solves.  Here one can instead use a stiff penalty; revisiting the multigrid approach via hierarchical coarsening~\cite{Botsch:2004:SGP,Shi:2006:TOG} may also prove fruitful.  Our approximation of tangent-point energy becomes inaccurate in situations of very tight contact (\ala{} \secrefs{ShrinkWrapping,NestedEnvelopes}), since we effectively have few quadrature points per unit surface area; adding additional quadrature points (or adaptive refinement) to elements in near-contact may help to achieve tighter fits.  For shape interpolation and mathematical visualization, it would be quite useful to find the trajectory that minimizes overall tangent-point energy, rather than just flowing to a common minimizer---here ideas about shell-space geodesics may prove valuable~\cite{Heeren:2012:TDG}.  Likewise, integrating repulsive regularization into a thin shell model might provide better proximity-aware shape editing by retaining a ``memory'' of the initial shape.  Finally, we do not directly treat boundary conditions, or more general arrangements of repulsive curves \emph{and} surfaces that might have interesting modeling applications.

\section*{Acknowledgments}

The authors thank Saul Schleimer and Henry Segerman for helpful discussions about topological examples.
This work was supported by a Packard Fellowship, NSF Award 1717320, and gifts from Autodesk, Activision Blizzard, Adobe, Disney, and Facebook.
The third author was supported by DFG-Project 282535003: \emph{Geometric curvature functionals: energy landscape and discrete methods}.

\bibliographystyle{ACM-Reference-Format}
\bibliography{RepulsiveSurfaces}

\appendix

\section{Action of the Fractional Operators}
\label{app:ActionOfTheFractionalLaplacian}

In section \secref{HierarchicalMatrices} we claimed that the actions of the fractional operators $L^\sigma$, $B$, and $B_0$ can be expressed by suitable kernel matrices that we then compress by hierarchical methods.
This is not obvious, so we include a brief derivation here.
Consider the kernel matrix
\[
\KernelMatrix_{ST} := (1-\delta_{ST}) \, 
	\nabs{\Bary{f}(S) - \Bary{f}(T)}^{-(2\sigma + 2)}.
\]
Rewriting \eqref{DiscreteLaplacianQuartic} for general $\mathbf{u}$ and $\mathbf{v} \in \RR^{\nabs{V}}$ in terms of this kernel yields
\[
	\textstyle
	\mathbf{u}^\T L^{\sigma} \mathbf{v} 
	= 
	\sum_{S \in F} \sum_{T \in F}
	(\FaceAv{u}(S) - \FaceAv{u}(T))
	\,
	(\FaceAv{v}(S) - \FaceAv{v}(T)) 
	\,
	a_f(S) \, \KernelMatrix_{ST} \, a_f(T).
\]
Multiplying the product inside the sum gives
\[
	\big( 
		\FaceAv{u}(S)\, \FaceAv{v}(S) 
		+ 
		\FaceAv{u}(T) \, \FaceAv{v}(T) 
		- 
		\FaceAv{u}(T) \, \FaceAv{v}(S) 
		- 
		\FaceAv{u}(S) \, \FaceAv{v}(T) 
	\big)  
	\,
	a_f(S) \, \KernelMatrix_{ST} \, a_f(T)
\]
for the pair $(S,T)$. Because $\KernelMatrix_{ST} = \KernelMatrix_{TS}$, we can move some terms between the summands for $(S,T)$ and $(T,S)$, and thus reorganize the sum into 
\begin{align*}
	\mathbf{u}^\T L^{\sigma} \mathbf{v} 
	&= 
	\textstyle	
	2 \, \sum_{S \in F} \sum_{T\in F}
	\big(
		\FaceAv{u}(S) \, \FaceAv{v}(S) 
		- 
		\FaceAv{u}(S) \, \FaceAv{v}(T) 
	\big) 
	\,
	a_f(S) \, \KernelMatrix_{ST} \, a_f(T) 
	\\
	&= 
	\textstyle
	2 \, \sum_{S \in F}
		\FaceAv{u}(S)  a_f(S)
		\left(
	    		a_f(S)^{-1} \sum_{T \in F} \KernelMatrix_{ST} a_f(T)
		\right)
		a_f(S) \FaceAv{v}(S) 		
	\\
	&\qquad
	-
	\textstyle
	 2\,
	\sum_{S \in F} \sum_{T\in F}
	\FaceAv{u}(S) \, a_f(S) \, \KernelMatrix_{ST} \, a_f(T) \, \FaceAv{v}(T).
\end{align*}
Recall that $\PreLo \in \operatorname{Hom}(\RR^{\nabs{V}}; \RR^{\nabs{F}})$ is defined by $(\PreLo \mathbf{u})(S) = a_f(S) \bar u(S)$. Thus the above collapses to
\begin{align*}
	\mathbf{u}^\T L^{\sigma} \mathbf{v} 
	&=
	2 \, \mathbf{u}^\T \PreLo^\T \diag(a_f)^{-1} \diag(\KernelMatrix \, a_f) \,\PreLo \,\mathbf{v}
	- 
	2\,
	\mathbf{u}^\T \PreLo^\T  \KernelMatrix \, \PreLo \, \mathbf{v}
	\\
	&=
	2\, 
	\mathbf{u}^\T  \PreLo^{\T}
	\big[\diag \big( \diag(a_f)^{-1} \KernelMatrix \, a_f \big) - \KernelMatrix \big] 
	\, 
	\PreLo
	\,
	\mathbf{v}
	.
\end{align*}
The derivation follows analogously for the high- and low-order matrices $B$ and $B_0$, with the substitution of the operator $\PreHi = \diag(a_f) \DMatrix$ for $\PreLo$ in the case of $B$.

\section{Fast Matrix-Vector Multiplication}
\label{app:FastMatrixVectorMultiplication}

Step~1 of \secref{HierarchicalMultiplication} corresponds to thinning out the matrix shown in \figref{BCT} by removing all the green parts. The remainder is a sparse block matrix with variable block size.
We store this sparse matrix in CSR format and perform matrix-vector multiplication via sparse BLAS routines.

In Step~2 the kernel matrix $\KernelMatrix_{\Ical\Jcal}$
is compressed into the rank-one-matrix 	
$\mathbf{1}_\Ical \,  
	\Kernel
		(X_{\Ical}, \Proj_{\Ical}  ;  
		X_{\Jcal}, \Proj_{\Jcal}
	) 
	\, 
	\mathbf{1}_\Jcal^{\T}
$.
In this step, we are cautious not to move the input data $\mathbf{x}_\Jcal$ and output data $\mathbf{y}_\Ical$ directly to and from the clusters $\Ical$ and $\Jcal$.
Instead, we employ a common technique for fast multipole and hierarchical matrix methods and
use the BVH for that. For each cluster $\Ical$, $\Jcal$, we allocate scalars $\tilde x_{\Jcal}$ and $\tilde y_{\Ical}$. We start only with the leaf clusters and set 
\[
	\tilde x_{\Jcal} \gets 
	\textstyle
	\sum_{T \in \Jcal} \mathbf{x}(T)
	\quad
	\text{for each leaf cluster $\Jcal$.}
\]
Then, during a parallel traversal of the BVH in post-order, for each cluster $\Jcal$, we add the $\tilde x$-values of its children into $\tilde x_{\Jcal}$.
After this \emph{upward pass} is finished, we loop over all clusters $\Ical$ and set
\begin{align}
\label{eq:FarFieldAction}
	\tilde y_{\Ical} 
	\gets 
	\textstyle
	\sum_{\Jcal}
	\Kernel
		(X_{\Ical}, \Proj_{\Ical}  ;  
		X_{\Jcal}, \Proj_{\Jcal}
	)  \, \tilde x_{\Jcal},
\end{align}
where the sum runs over the $\Jcal$ such that $(\Ical,\Jcal)$ is admissible.
This operation is also best performed by a sparse matrix multiplication. 
To this end, we fix an ordering of the BVH clusters, \eg, depth-first ordering.
Then we assemble a sparse matrix $\tilde \KernelMatrix$ 
with the nonzero value $\Kernel(X_{\Ical}, \Proj_{\Ical} ; X_{\Jcal}, \Proj_\Jcal )$ at the position that correspond to the admissible block cluster $(\Ical, \Jcal)$.
Storing $\tilde x$ and $\tilde y$ as vectors, \eqref{FarFieldAction} amounts to
\begin{align*}
	\tilde y \gets \tilde \KernelMatrix \, \tilde x.
\end{align*}
Afterwards, we use a \emph{downward pass} through the BVH to distribute the $\tilde y$-values back into the vector $\mathbf{y}$: We traverse the BVH in pre-order and let each cluster $\Ical$ add its $\tilde y$-value into each of its children's $\tilde y$-values.
Finally each leaf cluster adds its value into each of its member's $\mathbf{y}$-entry, \ie,
\begin{align*}
	\mathbf{y}(S) \gets \mathbf{y}(S) + \tilde y_{\Ical}
	\quad \text{for each leaf $\Ical$ and each $S \in \Ical$.}
\end{align*}

The structure of the kernel matrices of $L^\sigma$, $B$, and $B_0$ is very similar.
This allows us to use a single block cluster tree to compress all of them. Moreover, the sparsity patterns for the two sparse matrices used to perform Steps~1 and 2 can be shared and the corresponding nonzero values can be computed in a single parallelized loop over the admissible and inadmissible blocks, respectively.

For the application of $A_3$ to a vector $\mathbf{v}$ of size $3 \nabs{V}$, we could apply $A = B + B_0$ separately on three vectors $\mathbf{v}_1$, $\mathbf{v}_2$, and $\mathbf{v}_3$ of size $\nabs{V}$ that each store only one spatial component of the vertex positions.
However, it turns out to be more efficient to store $\mathbf{v}_1$, $\mathbf{v}_2$, and $\mathbf{v}_3$ as columns of a matrix of size $\nabs{V} \times 3$ and to replace the sparse matrix-vector products by sparse matrix-dense matrix products.

\end{document}